\documentclass{PoS}

\def\slash#1{#1\!\!\!\!\!/\!\,\,}
\def\Dslash{\slash D}

\newcommand{\be}{\begin{equation}}
\newcommand{\ee}{\end{equation}}
\newcommand{\bea}{\begin{eqnarray}}
\newcommand{\eea}{\end{eqnarray}}
\def\beann{\begin{eqnarray*}} 
\def\eeann{\end{eqnarray*}}

\title{
\vspace*{-2cm}
\begin{minipage}{\textwidth}
\begin{flushright}
\texttt{\footnotesize
PoS(LAT2007)178\\%
CERN-PH-TH/2007-186\\%
MS-TP-07-29\\
}
\end{flushright}
\end{minipage}\\[15pt]
\vspace*{+2cm}
\mbox{A QCD chiral critical point at small chemical potential:} \\
is it there or not?}

\ShortTitle{QCD critical point}

\author{\speaker{Philippe de Forcrand\footnote{Speaker}} \\
        Institut f\"ur Theoretische Physik, ETH Z\"urich, CH-8093 Z\"urich, Switzerland ~ and \\
        CERN, Physics Department, TH Unit, CH-1211 Geneva 23, Switzerland \\
        E-mail: \email{forcrand@phys.ethz.ch}}

\author{Seyong Kim\\
        Department of Physics, Sejong University, Seoul 143-747, Korea \\
        E-mail: \email{skim@sejong.ac.kr}}

\author{\speaker{Owe Philipsen\footnote{Speaker}} \\
        Institut f\"ur Theoretische Physik, Westf\"alische Wilhelms-Universit\"at M\"unster, Germany \\
        E-mail: \email{ophil@uni-muenster.de}}

\abstract{
For a QCD 
chiral critical point to exist, the parameter region of small quark masses for which 
the finite temperature transition is first-order must expand when the chemical
potential is turned on. This can be tested by a Taylor expansion of the
critical surface $(m_{u,d},m_s)_c(\mu)$. We present a new method to perform
this Taylor expansion numerically, which we first test on an effective 
model of QCD with static, dense quarks. We then present the results for
QCD with 3 degenerate flavors. For a lattice with $N_t=4$ time-slices,
the first-order region shrinks as the chemical potential is turned on.
This implies that,
for physical quark masses, the analytic crossover which occurs at $\mu=0$
between the hadronic and the plasma regimes remains crossover in the 
$\mu$-region where a Taylor expansion is reliable, i.e. $\mu \lesssim T$.
We present preliminary results from finer lattices indicating 
that this situation persists, as does the discrepancy between the 
curvature of $T_c(m_c(\mu=0),\mu)$ and the experimentally observed freeze-out curve.
}

\FullConference{The XXV International Symposium on Lattice Field Theory\\
		 July 30-4 August 2007\\
		 Regensburg, Germany}

\begin{document}

\section{Introduction}

The experimental determination of the QCD phase diagram is underway via extensive 
heavy-ion collision programs. At the same time, much effort is being devoted to its
theoretical determination via numerical lattice simulations. In the latter case,
one can ask the more general theoretical question: what is the phase diagram of 
$2+1$-flavor QCD in the $(m_{u,d},m_s,\mu_{u,d},\mu_s,T)$ parameter space?

\begin{figure}[t!]
\vspace*{-0.3cm}
\centerline{
\scalebox{0.40}{\includegraphics{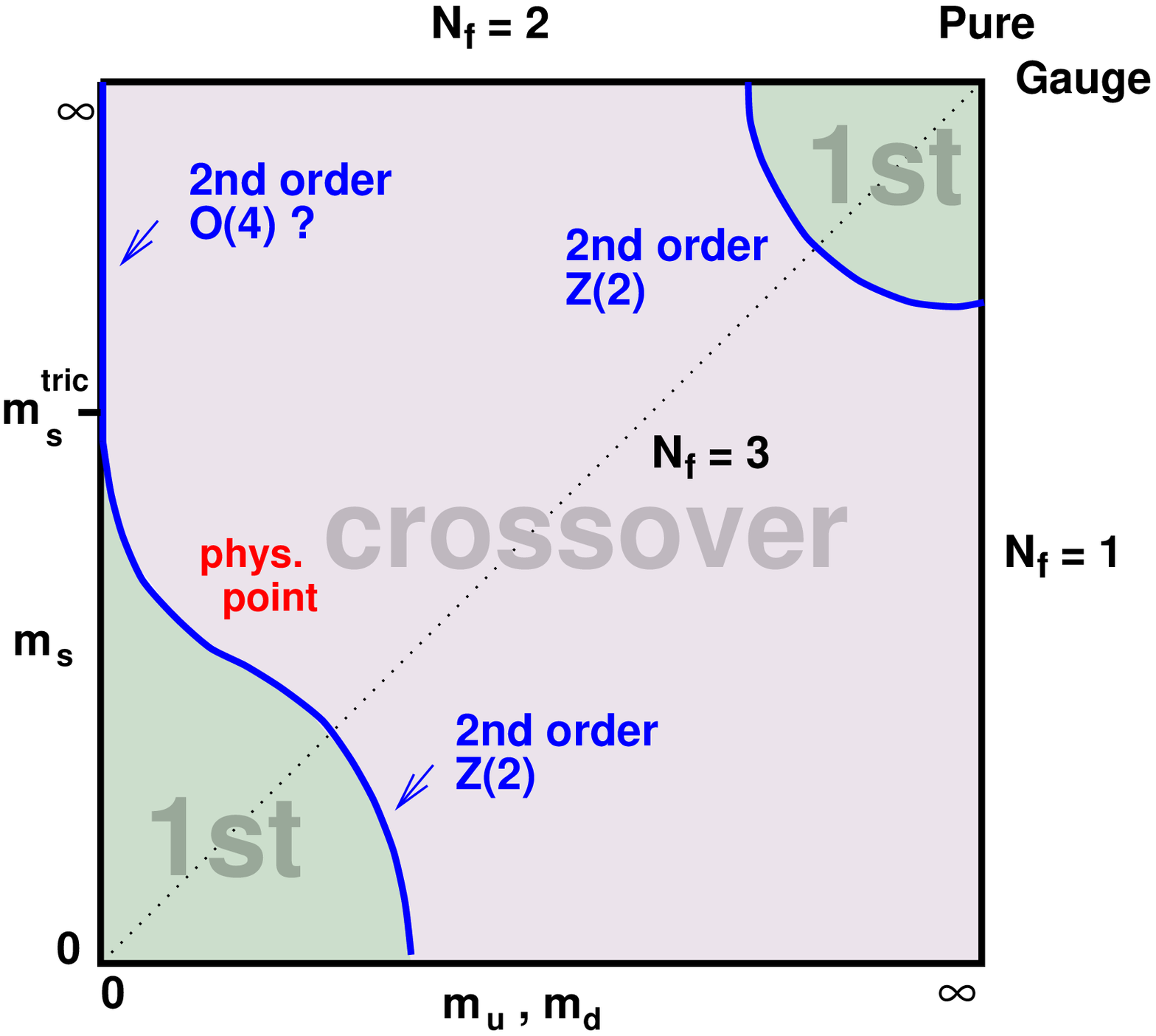}}
\hspace*{0.1cm}
\scalebox{0.63}{\includegraphics{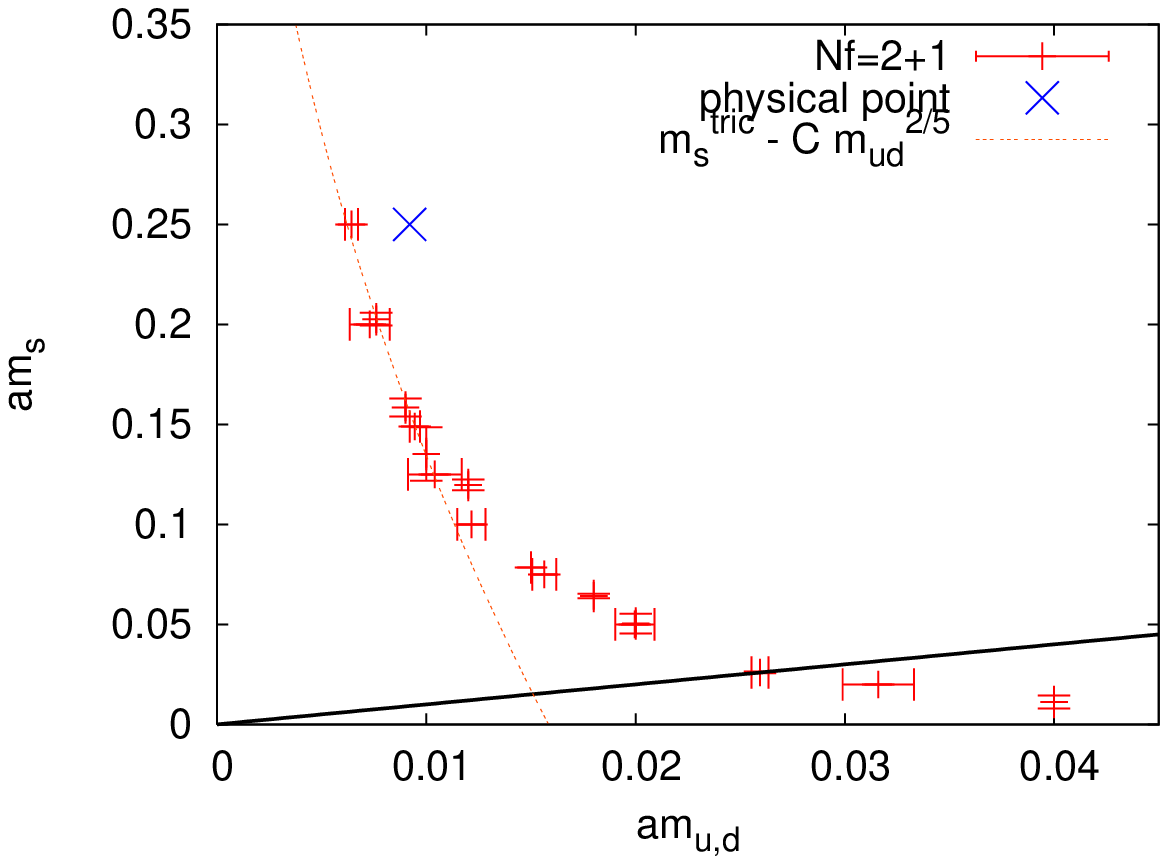}}
}
\caption{Schematic phase transition behaviour of $N_f=2+1$ flavor QCD for
different choices of quark masses $(m_{u,d},m_s)$ at $\mu=0$~\cite{OP_review} 
({\em left}), and numerical determination of the chiral critical line~\cite{JHEP} 
({\em right}).
}
\end{figure}

For $\mu_{u,d}=\mu_s=0$, theoretical expectations are summarized in Fig.~1~(left).
In the chiral ($m_{u,d}=m_s=0$) and the quenched ($m_{u,d}=m_s=\infty$) corners,
order parameters probing the breaking of the chiral and the center symmetries
exist, and the symmetry breaking or restoring transitions are first-order.
For intermediate quark masses, simulations show a crossover. This leads to the
existence of two lines of critical points, both in the $3d$ Ising universality
class, delimiting the first-order regions. \linebreak In the quenched corner, the critical
line has been studied in \cite{Feo,Dumitru}. In the chiral corner, Fig.~1~(right)
shows the result of \cite{JHEP}. Good agreement with expectations was found,
including consistency with a tricritical point 
($m_{u,d}=0,m_s=m_s^{\rm tric}$) with 
a rather heavy mass $m_s^{\rm tric} \sim 2.8 T_c$. The physical point, marked by an X,
lies in the crossover region. These results were obtained with standard staggered
fermions on an $N_t=4$ ($a\sim 0.3$ fm) lattice. 
One task is now to quantify cutoff effects and extrapolate to the continuum limit.
This extrapolation has been performed for the physical point~\cite{Nature}, 
confirming that it remains in the crossover region. Our preliminary results on 
an $N_t=6$ lattice,
consistent with those of \cite{Fodor_Nt6} and with earlier indications
from improved actions on $N_t=4$ 
\cite{Karsch_improved}, are presented Sec.~\ref{sec:Nt6}. Cutoff effects
are large, and the transition becomes much weaker on a finer lattice.

\begin{figure}[t!]
\vspace*{-0.9cm}
\centerline{
\scalebox{0.68}{\includegraphics{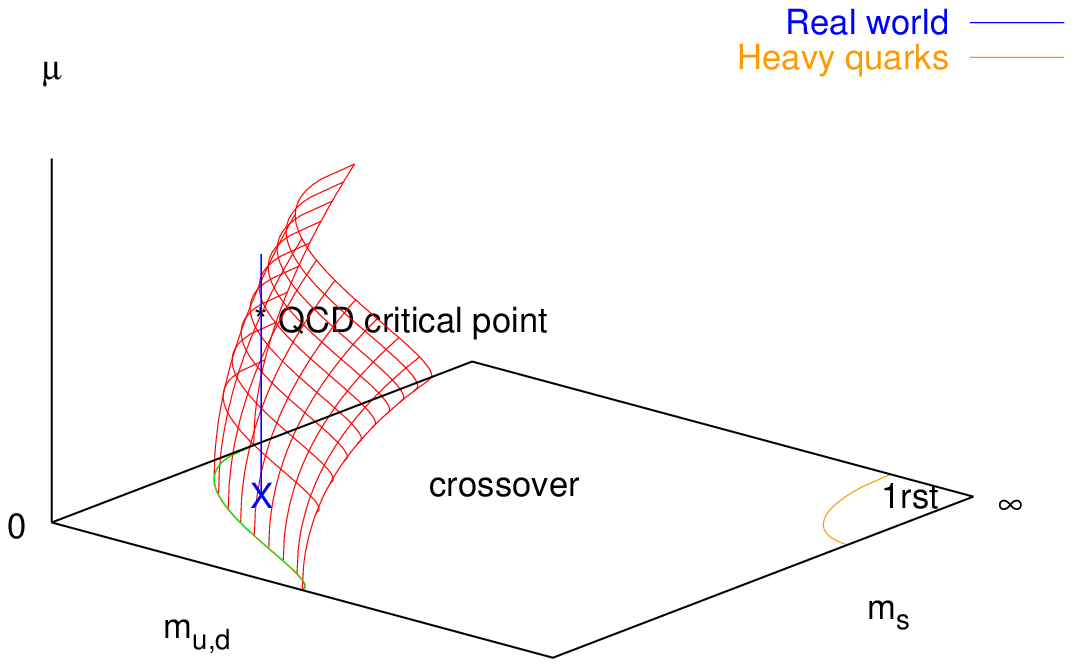}}
\hspace*{-0.5cm}
\scalebox{0.68}{\includegraphics{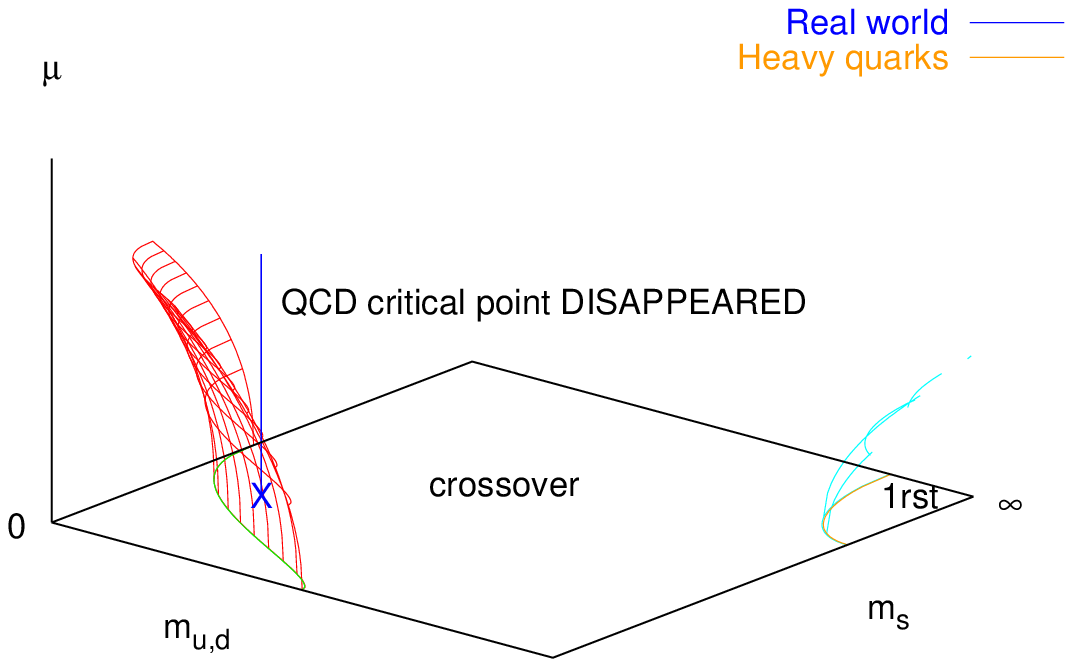}}
}
\caption{
For physical $(m_{u,d},m_s)$ quark masses, the finite-temperature ``transition''
at $\mu=0$ is really a crossover. As $\mu$ is turned on, this crossover 
becomes a genuine phase transition at the QCD chiral critical point ({\em left}),
provided the region of first-order transitions at small quark masses
{\em expands} with $\mu$. If not ({\em right}), there is no QCD chiral critical 
point. The curvature of the critical surface $dm_c/d\mu^2$ at $\mu=0$
distinguishes between these two possibilities for small $\mu$.
Note that, for heavy quarks, the first-order region {\em shrinks} when
$\mu$ is turned on~\cite{Potts}.
\label{fig2}
\vspace*{-0.3cm}
}
\end{figure}

The next issue is to determine the effect of a baryonic chemical potential.
For physical quark masses, it is expected that the transition becomes stronger 
with $\mu$, so that it turns from a crossover at $\mu=0$ into a first-order
transition, thus defining the QCD critical point $(\mu_E,T_E)$ where the transition 
is second-order. 
The fermion determinant becomes complex when $\mu\neq 0$, and the ensuing
``sign problem'' forbids standard Monte Carlo sampling.
Nevertheless, a determination of this critical point has been performed,
using the same staggered action and $N_t=4$ lattice spacing as us~\cite{FK_crit}.
While the cutoff error is likely to be comparable to that at $\mu=0$,
the numerical method used, namely reweighting,
introduces new possible systematic errors.
Contrary to the well-known multi-histogram reweighting where one {\em interpolates}
between Monte Carlo results obtained at several values of the coupling, here one
{\em extrapolates} results obtained at a single value of the coupling, $\mu=0$,
to $\mu\neq 0$.
While this procedure is in principle exact for infinite statistics, the practical 
question is whether the finite Monte Carlo sample contains the relevant information
at the extrapolated coupling. This difficulty is known as the overlap problem.
Thus, a crosscheck of the results of \cite{FK_crit} using a different
approach seems worthwhile.

To avoid all difficulties caused by a potential lack of overlap, we set a more
modest goal, and study the effect of an infinitesimal chemical potential. More
precisely, we consider 
the critical surface, swept by the chiral critical line of Fig.~1~(right) as a function of $\mu$,
in terms of a Taylor expansion in $(\mu/T)^2$ about $\mu=0$, 
(for the $N_f=3$ case to keep the notation simple)
\be
\frac{m_c(\mu)}{m_c(0)} = 1 + \sum_{k=1} c_k \left(\frac{\mu}{\pi T}\right)^{2k} \quad .
\label{eq:m_c}
\ee
The sign of the first Taylor coefficient $c_1$ is of crucial relevance to the
QCD critical point. Since for $\mu=0$ the physical point is in the crossover region,
the first-order region must expand with $\mu$ (Fig.~2~left). If instead the 
first-order region shrinks (Fig.~2~right), then the crossover persists and there
is no QCD critical point unless another critical surface, topologically unrelated
to the chiral critical surface we consider, is present in the phase diagram.
The simplest corresponding $(\mu,T)$ phase diagram in the two cases is depicted 
in Fig.~3, for a quark mass below and above $m_c(0)$. In the ``exotic'' scenario
(right), the first-order line present for small quark mass turns into a crossover
as the mass increases, 
starting with {\em large} $\mu$. If $m > m_c(0)$ as in the physical situation, the ``transition''
is simply a crossover, for all $\mu$'s, until different physics take over.
Distinguishing between these two scenarios requires full knowledge of the critical
surface. However, by measuring the first Taylor coefficient, we can determine the
behaviour of the critical surface for $\mu/T \lesssim 1$, i.e. in the region
$\mu_B \lesssim 500$ MeV where experimental searches are considered.

\begin{figure}[t]
\vspace*{-0.7cm}
\centerline{
\includegraphics[width=4.8cm,height=4.35cm]{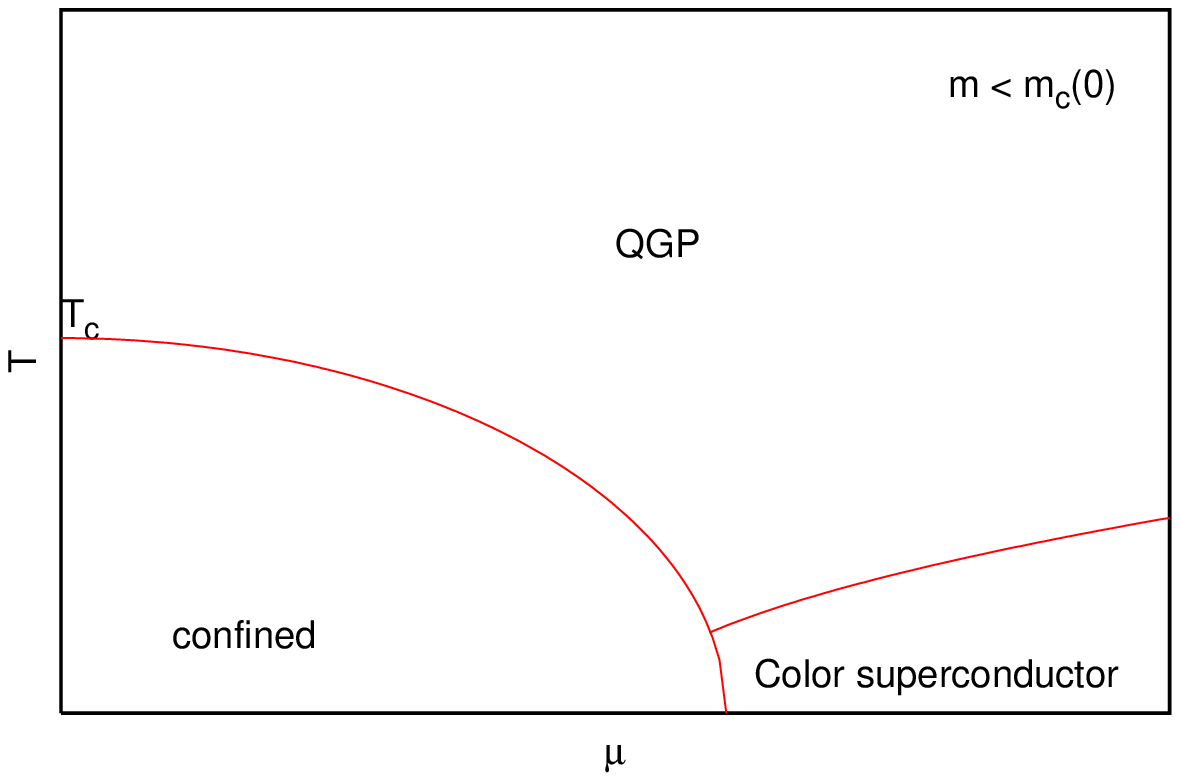}
\hspace*{0.2cm}
\includegraphics[width=4.5cm]{./s_ud_plane_lat99_2.eps}
\hspace*{0.2cm}
\includegraphics[width=4.8cm,height=4.00cm]{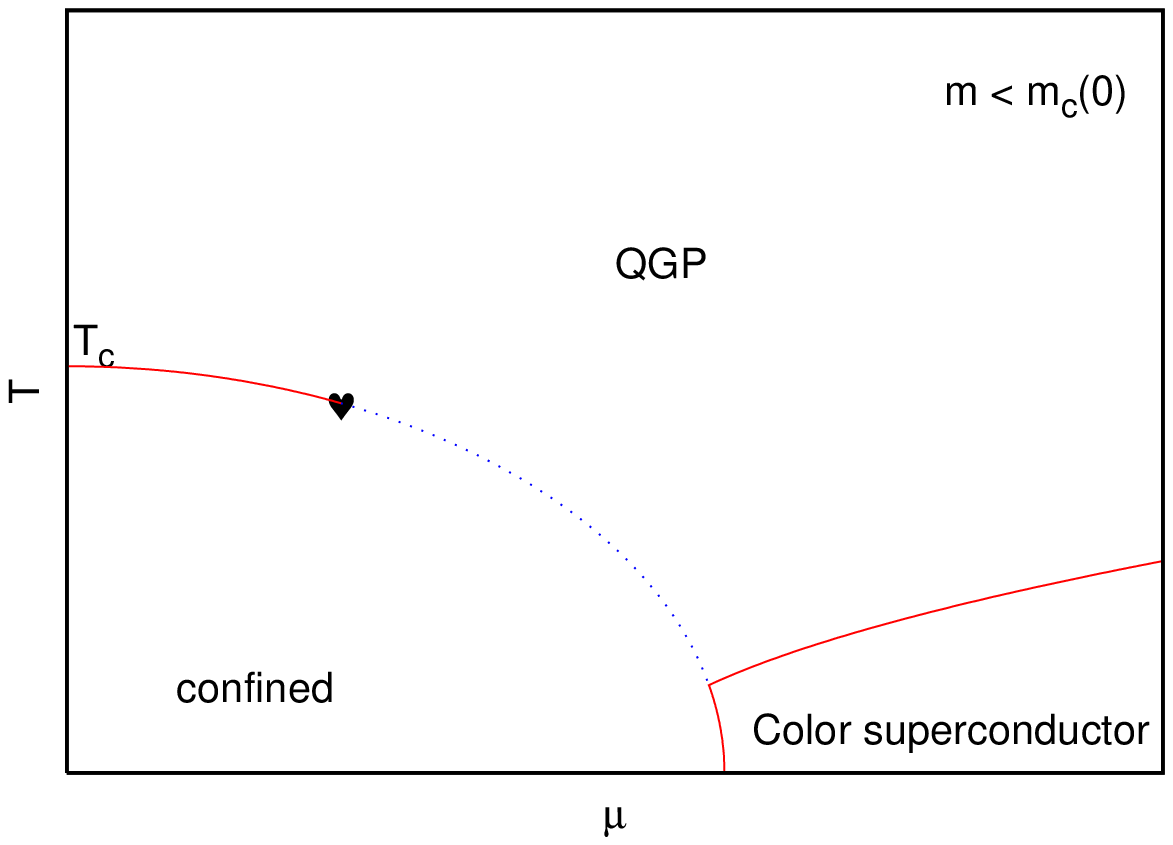}
\put(-247,26){$\bullet$}
}
\centerline{
\includegraphics[width=4.8cm,height=4.35cm]{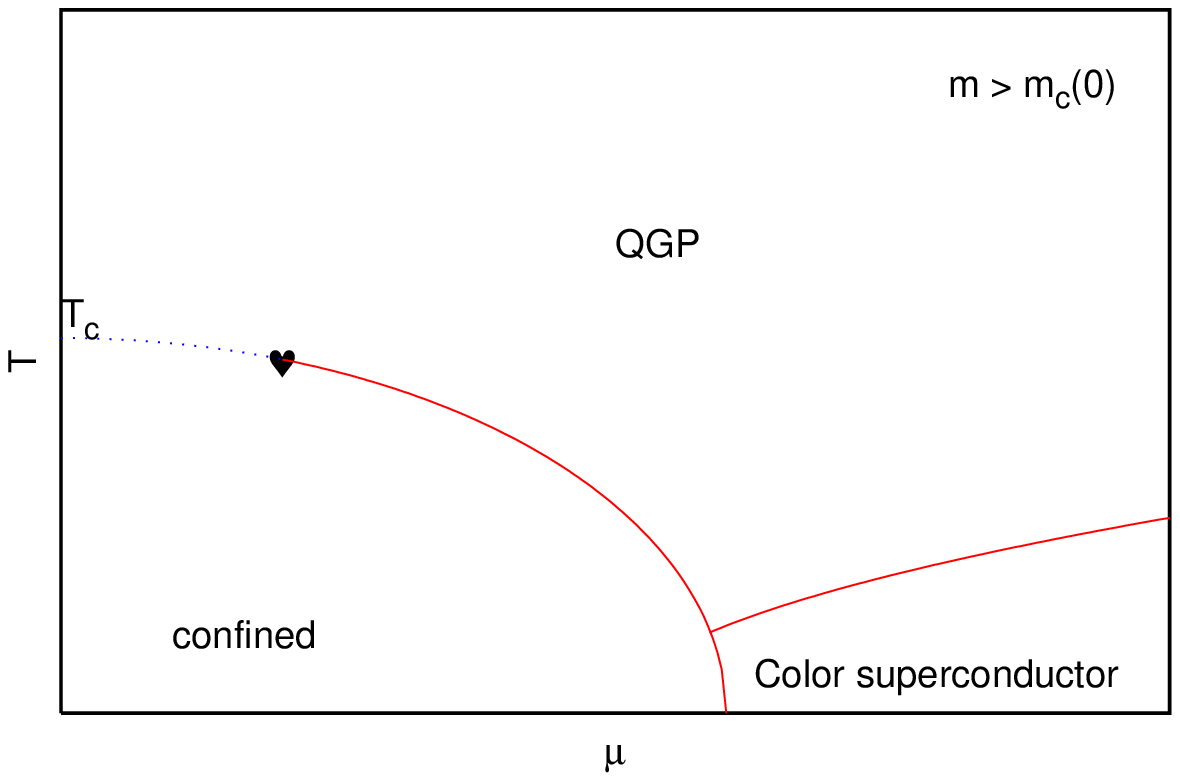}
\hspace*{0.2cm}
\includegraphics[width=4.5cm]{./s_ud_plane_lat99_2.eps}
\hspace*{0.2cm}
\includegraphics[width=4.8cm,height=4.00cm]{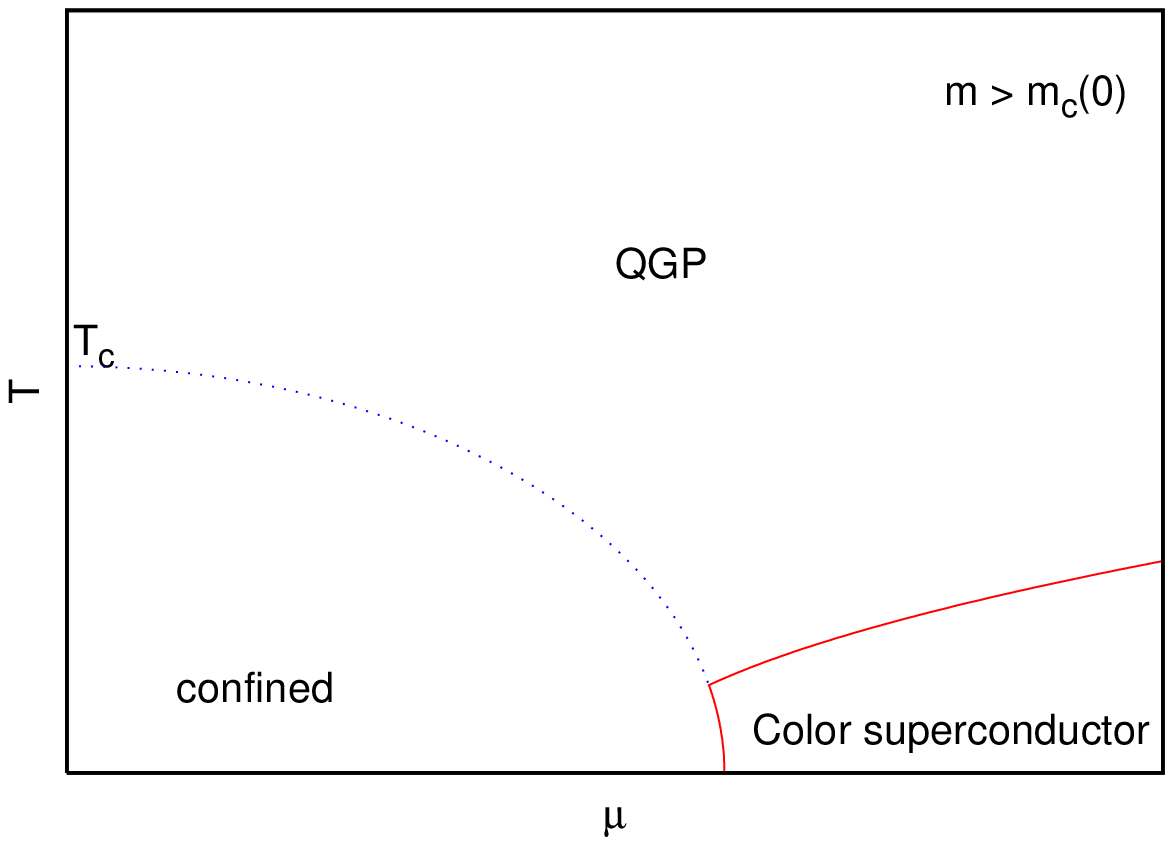}
\put(-237,36){$\bullet$}
}
\caption{
Simplest phase diagram in the $(\mu,T)$ plane, for the two cases
of Fig.~\protect\ref{fig2} 
({\em left} and {\em right}), as a function of the quark 
mass ({\em middle}). Red and blue lines indicate first-order transition
and crossover, respectively.
}
\end{figure}

In our first study~\cite{JHEP}, we performed simulations at imaginary $\mu$
and determined $m_c(\mu=i\mu_I)$, then fitted our results by a truncated Taylor
expansion which can be trivially continued to real $\mu$. 
While this method appears to work very well for the pseudo-critical line
$T_c(\mu)$~\cite{Nf2,Bielefeld_Swansea}, we faced two related difficulties when computing $m_c(\mu)$: the signal is 
very noisy,
and this makes the determination of the systematic error due to truncation of the
Taylor series difficult. 
For the case of three degenerate flavors, $m_s=m_{u,d}, \mu_s=\mu_{u,d}$, we
performed quadratic and quartic fits in $\mu/T$. The quartic term was statistically
insignificant, so we set it to zero and obtained for the leading term
\be
\frac{m_c(\mu)}{m_c(0)} = 1 - 0.7(4) \left( \frac{\mu}{\pi T} \right)^2 + \cdots
\ee
which favors the ``exotic'' scenario Fig.~2~(right), but not even at the $2\sigma$ level.
Note that including the quartic term in the fit (Table 2, line 2 of Ref.~\cite{JHEP})
changes the leading coefficient to -2.6(1.2).
The lack of compelling numerical evidence motivated us to improve our statistics and
our numerical methods.

Here, we compare three methods to obtain the derivative $\frac{dm_c}{d\mu^2}|_{\mu=0}$: \\
{\bf A}. Analytic continuation from imaginary $\mu$ \\
{\bf B}. Direct measurement of the derivative at $\mu=0$ \\
{\bf C}. Noisy reweighting to small imaginary $\mu$ \\
In all three methods, the same criterion is used to determine the pseudo-critical couplings
in a finite volume: the Binder cumulant 
$B_4 \equiv \frac{\langle (\delta X)^4 \rangle}{\langle (\delta X)^2 \rangle^2}$,
with $\delta X = X - \langle X \rangle$ and $X = \bar\psi \psi$,
takes the value 1.604 characteristic of the $3d$ Ising universality class
at the critical point. In a finite volume, $B_4$ is an analytic function of
$(m,T,\mu)$, and $T$ is fixed to the pseudo-critical temperature $T_c(m,\mu)$ by requiring,
for instance,
that the susceptibility $\langle (\delta X)^2 \rangle$ be maximum. We use the equivalent
prescription $\langle (\delta X)^3 \rangle = 0$. 
The two constraints $\{B_4=1.604, \langle (\delta\bar\psi \psi )^3 \rangle =0\}$ define a line in the $(T,m,\mu)$ space. 
We want to extract the curvature $\frac{dm_c}{d\mu^2}$ of this line at $\mu=0$.

Method A is the one used in \cite{JHEP}. The results of different simulations 
at various imaginary values of $\mu$ are fitted with a truncated Taylor expansion
about $\mu=0$. One difficulty comes from the systematic error associated with the
order of the truncation. Another comes from the final statistical error, which
in \cite{JHEP} amounted to about half the signal. 

Method B is reminiscent of the Bielefeld-Swansea approach \cite{Bielefeld_Swansea},
where the derivatives of the free energy with respect to $\mu/T$ are expressed as
expectation values of non-local operators, which involve traces of inverse powers
of the Dirac operator. In our case, the derivatives of $B_4$ and $\langle (\delta\bar\psi \psi )^3 \rangle$ with respect
to $m, T$ and $\mu$ can be expressed as expectation values of complicated operators,
which can be measured in a single simulation at $\mu=0$. 

Method C, to our knowledge, is a new application of reweighting, where the reweighting
factors are not 
evaluated exactly but estimated by a stochastic method. This noise does
not prevent reweighting provided it is unbiased. We can apply this strategy to the
results of $\mu=0$ simulations and reweight them to imaginary $\mu$, monitoring
the change in $B_4$ and $\langle (\delta\bar\psi \psi )^3 \rangle$. 
Keeping this imaginary $\mu$ very small has two advantages: $(i)$ the reweighting factor 
remains close to 1 and the overlap of the $\mu=0$ Monte Carlo ensemble with the 
target $\mu\neq 0$ ensemble is guaranteed; $(ii)$ fluctuations in the two ensembles 
are strongly correlated, and cancel in the observables.

In Section 2,
we first test methods B and C on a Potts model representative of the heavy-quark
limit of QCD, for which we have previously determined the critical line for both
imaginary and real $\mu$~\cite{Potts}. 
All methods agree, with a clear advantage to method C for its simplicity.
Then in Section 3, we compare methods A and C for $N_f=3$ QCD, on a coarse
$8^3\times 4$ lattice. 
Again, methods A and C agree, leaving no doubt that for this system the first-order
region {\em shrinks} when the chemical potential is turned on.
Finally, in Section 4 we present some preliminary results on a finer 
$18^3\times 6$ lattice, and discuss the implications and limitations of our findings.


\section{Potts model}

\subsection{An effective description of dense static quarks}

When the quark mass $m$ becomes infinite, quarks become static and 
a quark source is a Polyakov loop. The canonical partition function of QCD with
$n$ static quarks and $\tilde n$ static antiquarks is~\cite{Chandra}
\be
Z_{n,\tilde n} = \int {\cal D}A \frac{1}{n!} \Phi[A]^n \frac{1}{{\tilde n}!} \Phi^*[A]^{\tilde n} \exp(-(n+\tilde n)\frac{m}{T}) \exp(-S_g[A]) \quad ,
\ee
where $\Phi[A]$ is the 
Polyakov loop integrated over space and $S_g$ the gauge action. 
The grand-canonical partition function is then simply
\be
Z(\mu) = \sum_{n,\tilde n} e^{(n-\tilde n) \frac{\mu}{T}} Z_{n,\tilde n} 
= \int {\cal D}A \exp(-S_g[A]+ e^{-\frac{m-\mu}{T}}\Phi[A] + e^{-\frac{m+\mu}{T}}\Phi^*[A]) \quad .
\ee
Simplifying the gauge action to a Potts interaction between neighbouring Polyakov loops, 
one obtains the partition function of a $3d$ $q=3$ Potts model in an external field:
\be
Z(\kappa,\bar m,\bar \mu) = \sum_{\{\Phi(\vec{x})\}} \exp [-\kappa \sum_{i,\vec{x}} \delta_{\Phi 
    (\vec{x}), \Phi (\vec{x}+i)} + \sum_{\vec{x}} (h \Phi (\vec{x}) +
  h' \Phi^* (\vec{x}) )] \quad ,
\label{z3potts} 
\ee
where
$\Phi(\vec x)$ is the Potts (actually $Z_3$) spin $\exp(i k(\vec x) \frac{2\pi}{3})$, 
$h = h_m e^{+\bar\mu}, h' = h_m e^{-\bar\mu}, h_m=e^{-\bar m}, \bar\mu=\frac{\mu}{T}$
and $\bar m = \frac{m}{T}$.
Contrary to \cite{Chandra} who consider the limit $\bar m, \bar\mu \to \infty, (\bar m -\bar\mu)$ finite, and drop the last term corresponding to antiquarks, we keep the
complete expression in order to study the $(\bar m,\bar\mu)$ phase diagram.
Because $h' \neq h^*$ when $\mu\neq 0$, the action becomes complex and a sign
problem appears. However, it is very mild, so that simulating with the real part of
the action and reweighting with the imaginary part works efficiently and reliably. 
Two phases are present:
confined/disordered at small $\kappa$, deconfined/ordered at large
$\kappa$. The phase transition is first-order for the ordinary Potts
model, i.e., in the absence of any external field $h=h'=0$. This case corresponds
to $\bar{m} \to \infty$. 

\begin{figure}[t!]
\centerline{
\scalebox{0.32}{\includegraphics{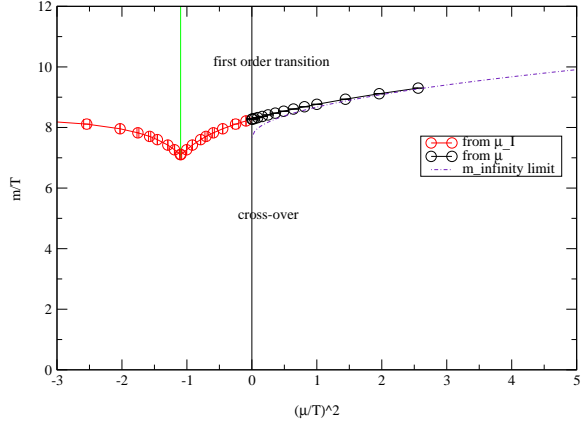}}
}
\caption{
Phase diagram of the Potts model, as a function of $m/T$ and $(\mu/T)^2$. 
For heavy masses, the transition is first order. It turns into a crossover 
for light masses. The critical line can be determined for imaginary $\mu$
(red points), showing the expected $Z_3$ transition at $\mu/T = i \frac{\pi}{3}$,
or directly for real $\mu$, because the sign problem is mild.
The line is analytic at $\mu=0$, but shows some curvature.
Methods B and C measure its slope at $\mu=0$.
}
\label{fig:Potts}
\end{figure}

Decreasing $\bar m$ turns on the magnetic field. This weakens the phase transition, 
so that it becomes 
a crossover for sufficiently large $h_m$, i.e., sufficiently small $\bar{m}=m/T$.
The critical value of $\bar{m}$ corresponding to the end of the first-order region
depends on the chemical potential $\bar\mu$. The critical line $\bar{m}_c(\bar\mu)$
has been determined in~\cite{Potts}, for both real and imaginary chemical potentials.
Fig.~\ref{fig:Potts} summarizes the results of \cite{Potts}. 
It represents the qualitative behaviour of QCD with heavy quarks near the critical
line in the upper right corner of Fig.~1 (left), because the symmetries and the infrared \linebreak
degrees of freedom are the same. Given $T_c\approx 270$ MeV, the value of $m_c(0)$ 
even lies in the 1-2 GeV range estimated from full QCD simulations~\cite{Feo,Dumitru}. 
Fig.~4 shows that, as the chemical potential is turned on, the first order region 
shrinks as indicated in the quenched corner of Fig.~2 (right).

As a function of 
${\bar\mu}^2$, the critical line is analytic at $\bar\mu=0$, so that the real-$\bar\mu$
dependence can be reconstructed by analytic continuation of imaginary-$\bar\mu$
simulation results. However, some curvature is visible, particularly as the
imaginary $\bar\mu$ approaches the $Z_3$ transition point $\bar\mu = i \frac{\pi}{3}$.
A global fit of all our data requires a fourth-order polynomial in ${\bar\mu}^2$:
\be
\frac{m}{T} = 8.273 + 0.585 \left(\frac{\mu}{T}\right)^2 - 0.174 \left(\frac{\mu}{T}\right)^4
+ 0.160 \left(\frac{\mu}{T}\right)^6 -0.071 \left(\frac{\mu}{T}\right)^8 \quad ,
\ee
where the first coefficient is $0.585(3)$ including its fitting error.
This is essentially the result of method A (although we have actually augmented 
the imaginary-$\mu$ data with real-$\mu$ data).
One may be concerned that a similar curvature occurs in QCD as well, which would
make it difficult to analytically continue data from a few discrete values of
imaginary $\mu$.
Therefore, the object of this study of the Potts model 
is to see how well the first coefficient
of the polynomial, 0.585(3), can be reproduced by methods B and C which involve
simulations at $\mu=0$ only.

\subsection{Method B}

The order parameter probing the $Z_3$ symmetry is the magnetization 
$M \equiv \sum_{\vec x} \Phi(x)$.
Therefore, we determine the critical line by requiring
\be
B_4 \equiv \frac{\langle (\delta M)^4 \rangle}{\langle (\delta M)^2\rangle^2} = 1.604, \qquad
\langle (\delta M)^3 \rangle = 0 \quad ,
\label{eq:crit}
\ee
with $\delta M \equiv M - \langle M \rangle$.
Both expressions are functions of $(\kappa,\bar m,\bar\mu)$, so that the two
equations determine the critical line $\bar m_c(\bar\mu)$ in this 3-parameter space.
The two equations can be Taylor expanded about the point 
$(\kappa_c(\bar\mu=0), m_c(\bar\mu=0),0)$ of this line, yielding
at lowest order
\be
dB_4 = A d\bar m + B d{\bar\mu}^2 + C d\kappa, \qquad     
d\langle (\delta M)^3 \rangle = A' d\bar m + B' d{\bar\mu}^2 + C' d\kappa \quad .
\ee
Staying on the critical line implies that the changes in $B_4$ and in
$\langle (\delta M)^3 \rangle$ are both zero. From the resulting two linear equations,
one can eliminate $d\kappa$, and obtain
\be
\frac{d\bar m}{{d\bar\mu}^2} = - \frac{BC' - B'C}{AC' - A'C} \quad ,
\label{eq:slope}
\ee
which is the desired slope of the critical line at $\mu=0$. 
The coefficients $A,B,C,A',B',C'$ are the following expectation values
\bea
A &=& \frac{\partial B_4}{\partial \bar m}|_{\bar\mu=0}
   =  - 2 h_m \langle \delta M^2\rangle^{-2} \langle \delta M^5\rangle \nonumber \\
B &=& \frac{1}{2} \frac{\partial^2 B_4}{\partial {\bar\mu}^2}|_{\bar\mu=0} 
 = \frac{1}{2} \langle \delta M^2 \rangle^{-2} [ 2 h_m \langle
\delta M^5 \rangle + h_m^2 \langle (\delta M^4 - \langle \delta M^4
\rangle) (\Phi - \Phi^*)^2 \rangle ] 
\nonumber \\
& & \hspace*{3.0cm} - h_m^2 \langle \delta M^2\rangle^{-3} \langle \delta M^4\rangle \langle
(\delta M^2 - \langle \delta M^2 \rangle)(\Phi - \Phi^*)^2\rangle \nonumber \\
C &=& \frac{\partial B_4}{\partial \kappa}|_{\bar\mu=0}
   =  \langle \delta M^2\rangle^{-2} \langle \delta M^4 \delta E \rangle -
2 \langle \delta M^2 \rangle^{-3} \langle \delta M^4\rangle \langle
\delta M^2 \delta E \rangle \nonumber \\
A' &=& \frac{\partial \langle \delta M^3\rangle}{\partial \bar m}|_{\bar\mu=0}  
   =  - 2 h_m \langle \delta M^4 \rangle \nonumber \\
B' &=& \frac{1}{2} \frac{\partial^2 \langle \delta M^3
  \rangle}{\partial {\bar\mu}^2}|_{\bar\mu = 0} 
  =  \frac{1}{2} [ 2 h_m \langle (\delta M^3 - 3 \langle \delta M^2
  \rangle \delta M) M \rangle + h_m^2 \langle (\delta M^3 - 3 \langle
  \delta M^2 \rangle \delta M) (\Phi - \Phi^*)^2 \rangle ] \nonumber \\
C' &=& \frac{\partial \langle \delta M^3\rangle}{\partial \kappa}|_{\bar\mu=0}
   =  \langle \delta M^3 \delta E \rangle - 3 \langle \delta M^2
  \rangle \langle \delta M \delta E \rangle
\label{eq:coeffsB}
\eea
where $\delta E = E - \langle E \rangle$.
Thus, all 6 coefficients can be measured during a Monte Carlo simulation at 
the $\mu=0$ critical point, i.e., where $\kappa$ and $\bar m$ have been tuned to satisfy 
eq.(\ref{eq:crit}).
We have performed such measurements on a $72^3$ lattice, of the same size used to
obtain Fig.~4. Substituting into eq.(\ref{eq:slope}), and performing a jackknife
bin analysis, we obtain 
$d\bar m/d{\bar\mu}^2 = 0.593(8)$,
which agrees very well with the earlier result from method A. 
Note again that method B is insensitive to the curvature of the critical line,
unlike method A.

\subsection{Method C}

The previous method enforced algebraically that the Binder cumulant $B_4$ stay 
constant under an infinitesimal change $d\bar m$, $d{\bar\mu}^2$. 
Instead, one can measure the change in $B_4$, $\Delta B_4$, under a 
small variation $\Delta\bar m$ or $\Delta{\bar\mu}^2$,
thus estimating the finite differences
$\Delta B_4 / \Delta\bar m$ and $\Delta B_4 / \Delta{\bar\mu}^2$. 
For sufficiently small variations, these discrete differences will approach
the derivatives $\partial B_4 / \partial\bar m$ and $\partial B_4 / \partial{\bar\mu}^2$.
Finally, once the gradient of $B_4$ is known, the direction which keeps
$B_4$ constant is given by
\be
\lim_{\Delta\bar\mu \to 0} \frac{\Delta\bar m}{{\Delta\bar\mu}^2} = 
- \frac{\partial B_4}{\partial{\bar\mu}^2} / \frac{\partial B_4}{\partial \bar m} \quad .
\label{eq:slopeC}
\ee
The second equation in (\ref{eq:crit}) is not needed here. The measurement of $B_4$
at the shifted couplings $(\bar m_c + \Delta\bar m,\mu=0)$ or $(\bar m_c,\Delta\bar\mu)$
is performed by the usual analysis, which automatically tunes $\kappa$ to satisfy
$\langle (\delta M)^3 \rangle = 0$.

\begin{figure}[t!]
\vspace*{-0.3cm}
\centerline{
\scalebox{0.54}{\includegraphics{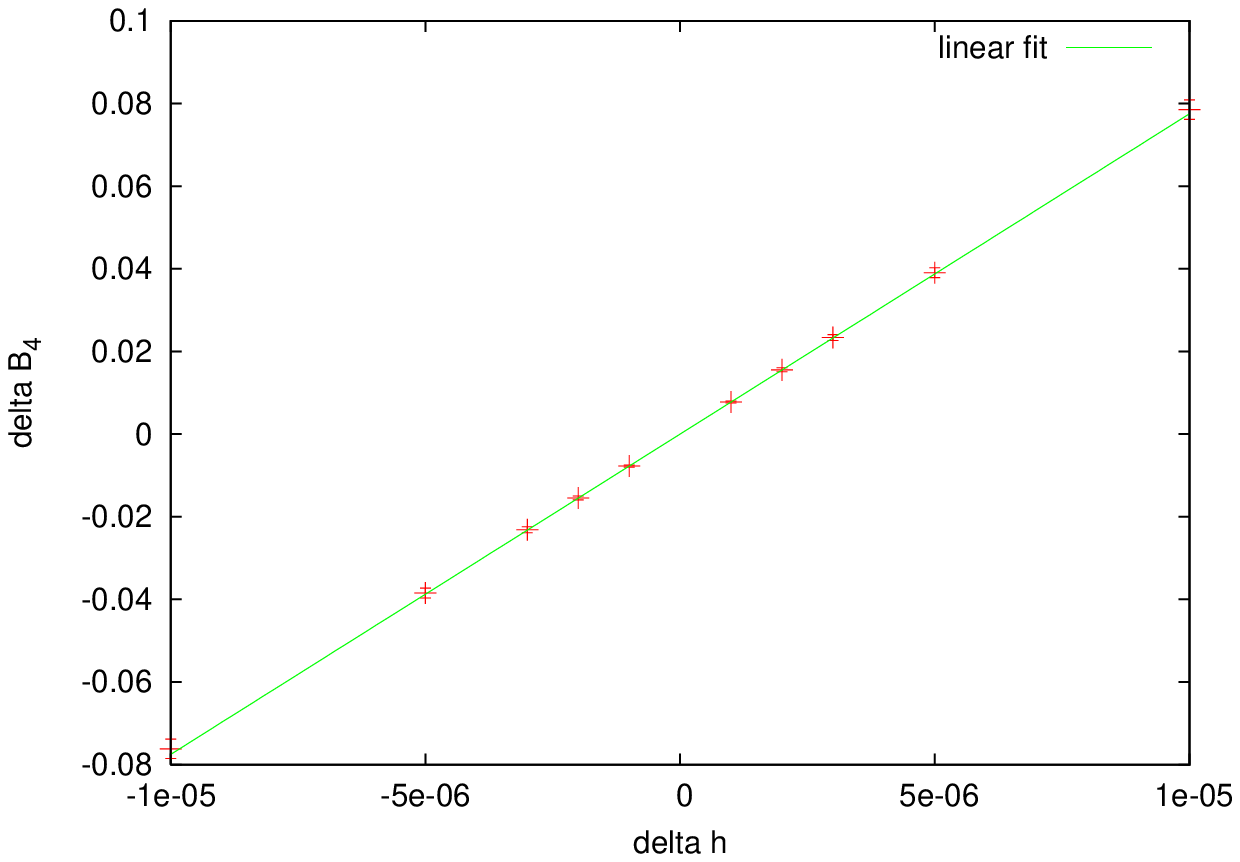}}
\hspace*{0.6cm}
\scalebox{0.54}{\includegraphics{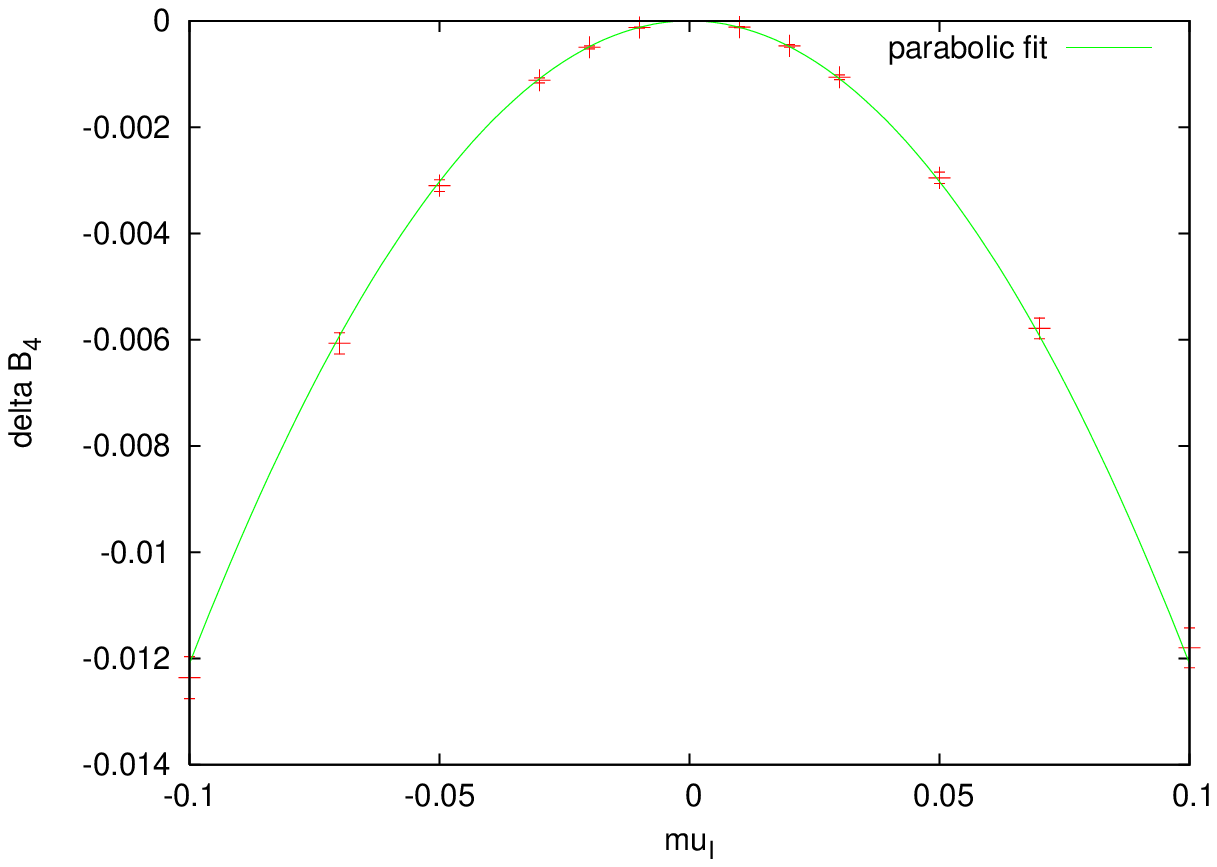}}
}
\centerline{
\scalebox{0.54}{\includegraphics{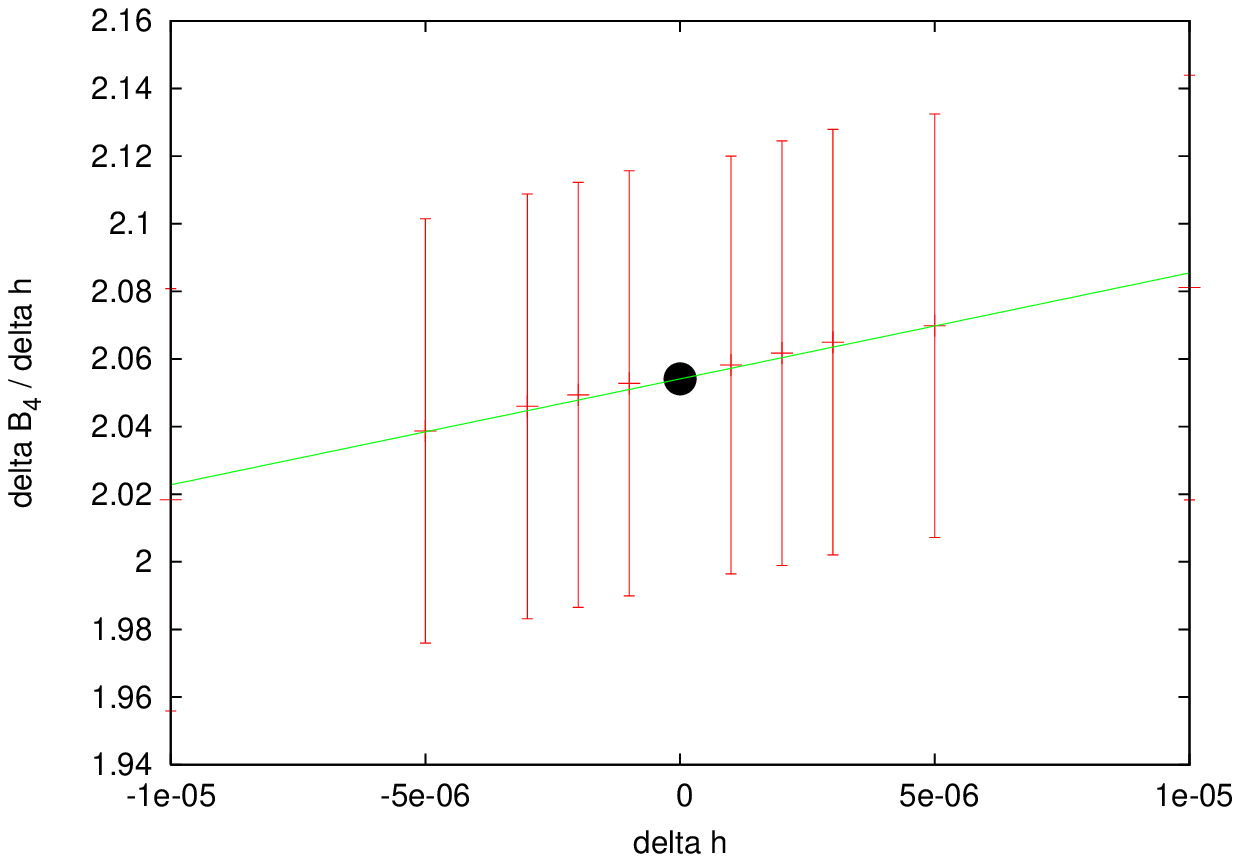}}
\hspace*{0.6cm}
\scalebox{0.54}{\includegraphics{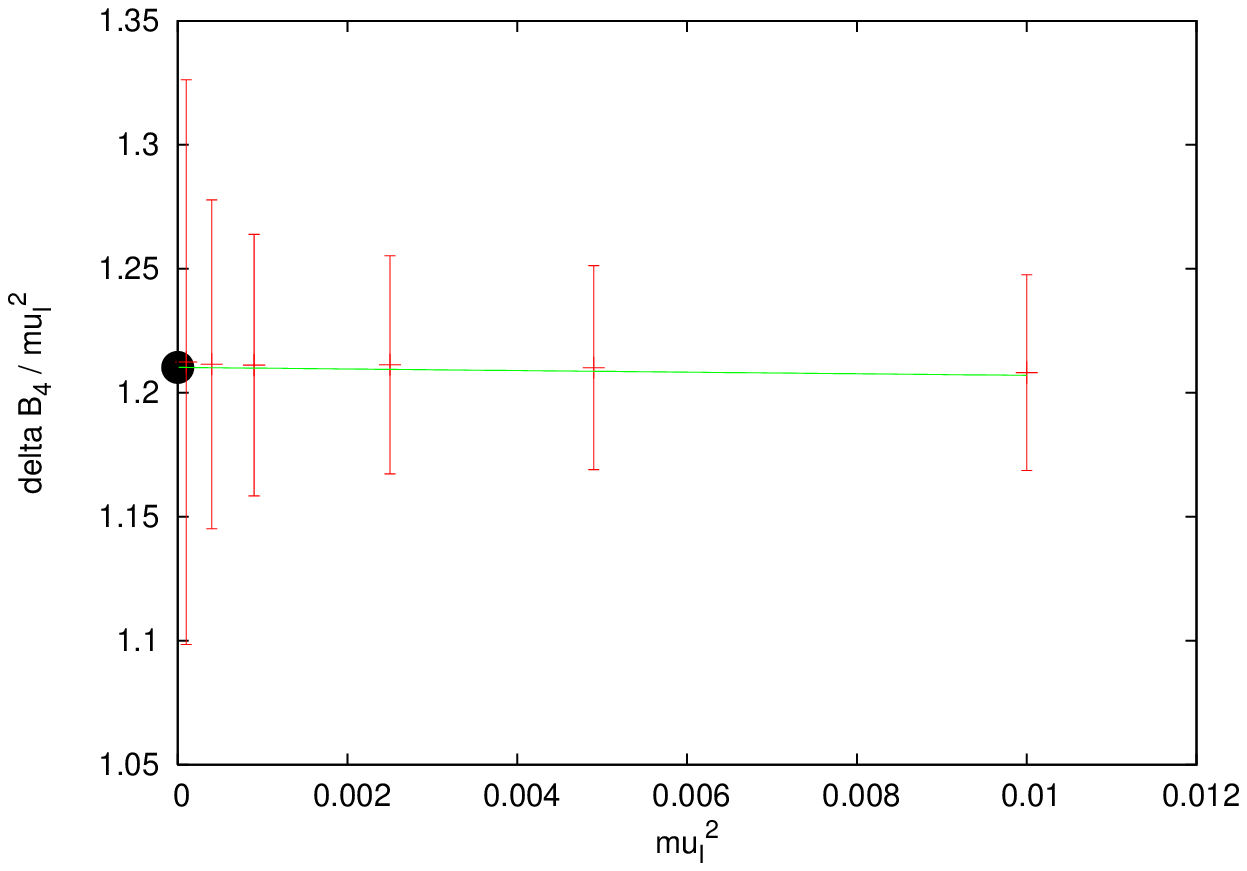}}
}
\caption{
Effect of a small change in the magnitude ({\em left}) and the orientation
({\em right}) of the magnetic field $h = h_m \exp(i\bar\mu_I)$ 
on the Binder cumulant $B_4(Re(M))$ in the Potts system.
}
\end{figure}

To measure $B_4$ at the shifted couplings, one could perform new Monte Carlo simulations.
But this is not necessary. Because the shift in the couplings is very small,
it is adequate and safe to use the original Monte Carlo ensemble and simply reweight
the results in the standard way~\cite{reweight}. Moreover, to avoid complex weights
one reweights to imaginary $\bar\mu$, which simply introduces a sign flip in 
eq.(\ref{eq:slopeC}). The reweighting factors remain real positive, and close to 1.

With reweighting, the fluctuations in the original and the
reweighted ensembles are strongly correlated. This can be turned into a virtue,
as these correlated fluctuations drop out
of our observable, which is the {\em change} in $B_4$, rather than $B_4$ itself.

This procedure is illustrated Fig.~5. The top row shows the change in $B_4$ under
a change in $h_m=\exp(-\bar m)$ (left), and under a change in imaginary $\bar\mu$ (right).
One observes the expected dependence, respectively linear and quadratic.
Note that the changes in $B_4$ are small, ${\cal O}(10^{-2})$, and measured
very accurately -- much more accurately than $B_4$ itself.
The bottom row of Fig.~5 shows the change in $B_4$, divided by the change in $h_m$
(left) or by $(\bar\mu)^2$ (right). If the Taylor expansion could be truncated to
leading order, the data would be constant. Instead, one sees the small influence
of the next Taylor order. A fit to (constant + linear) gives the desired partial
derivatives, marked by a black circle, which can be substituted in eq.(\ref{eq:slopeC}). 
The resulting slope is 0.589(7), again consistent with the other two methods.

Note that the statistical errors on the various points Fig.~5 are extremely correlated
with each other, so that a jackknife bin analysis is required to obtain reliable
errors on the final slope. Note also that there is a broad optimum for the shift
in the couplings: too small a shift produces too small a change in $B_4$, and the
estimates of the derivatives approach $0/0$; too large a shift introduces a systematic
error from higher-order Taylor terms and a potential overlap problem.

The errors from methods B and C are similar. This is normal,
since these two methods make use of the same Monte Carlo data and extract the
same information. So the preference
for method C in this case comes from its simplicity: there is no need to measure
the relatively complicated observables eqs.(\ref{eq:coeffsB}). 
This will become a more serious issue in the case of QCD.


\section{$N_f=3$ QCD}

In \cite{JHEP},
for $N_f=3$ QCD, with standard staggered fermions on an $N_t=4$ lattice, we
determined the critical quark mass at $\mu=0$: $a m^c_0= 0.0263(3)$,
and proceeded to Taylor expand the pseudo-critical coupling $\beta_c$ and
the Binder cumulant of $\bar\psi \psi$:
\bea
\label{eq:betac_QCD}
\beta_c(a\mu,am) & = & \sum_{k,l=0} c_{kl}\, (a\mu)^{2k}\, (am-am^c_0)^l \\
B_4(am,a\mu)&=&1.604+b_{10}\left[am-am^c_0-c_1'(a\mu)^2\right]+b_{20}(am-am^c_0)^2\nonumber\\
&&-b_{10}\left[(c_2'-c_1'C)(a\mu)^4+C(am-am^c_0)(a\mu)^2\right] + \cdots \quad ,
\label{eq:B4_QCD}
\eea
from which one can extract the variation of the pseudo-critical temperature and
of the critical quark mass with the chemical potential.

\subsection{Recall: method A}

In \cite{JHEP}, we performed independent simulations at different values of the
quark mass and imaginary chemical potential $\mu=i\mu_I$, which we then fitted with
the Taylor expansions above. 

For the pseudo-critical temperature, a leading order fit was satisfactory, yielding
for the $\mu^2$ dependence $c_{10}=0.781(7)$. At subleading order, there was no
evidence for a cross-term $(am - am^c_0) (a\mu)^2$, and the $(a\mu)^4$ term
was barely statistically significant. Including it in the fit yielded 
$c_{10}=0.759(22)$ (see Table 1 of \cite{JHEP}). 
Both fits are shown in Fig.~6 (left), by the narrow blue and 
broader green error band, respectively. 

For the curvature of the critical surface, 
the coefficient $c_1'$ eq.(\ref{eq:B4_QCD}) encodes the relevant information: 
to keep $B_4$ constant, one must satisfy $\frac{d(am)}{d(a\mu)^2} = c_1'$.
Again, no cross-term $C$ was visible, and 
fits including $\mu^2$ only, or $\mu^2$ and $\mu^4$ terms were performed,
resulting in the narrow blue and broader green error bands Fig.~6 (right),
corresponding to Table 2, lines 2 and 3 of \cite{JHEP}.
They suggest a negative value for $c_1'$, but do hardly more than that.

\subsection{Method B}

The Taylor coefficients in eqs.(\ref{eq:betac_QCD},\ref{eq:B4_QCD}) 
can be expressed as expectation
values to be measured at $\mu=0$. We wrote down these expressions, analogous to
eqs.(\ref{eq:coeffsB}). But we did not implement a program to measure these
operators, for several reasons. 
First, the programming effort is non-trivial. For instance, the trace of the $5^{th}$
inverse power of the Dirac operator must be evaluated to obtain the derivative 
of $(\delta\bar\psi \psi)^4$ with respect to the quark mass. 
Moreover, important cancellations will take place among the various contributions, leading 
to further difficulties with optimizing the number of noise vectors to be used as
stochastic estimators of the various traces. 
Finally, we realized, from the Potts test case, that method C makes use of
the same information contained in the $\mu=0$ Monte Carlo ensemble, and gives the
same output as method B with less effort.

\subsection{Method C}

As in the Potts case, one can shift very slightly the quark mass and the 
chemical potential, and reweight the $\mu=0$ Monte Carlo ensembles to these
shifted couplings. The effect of a shift in the quark mass was already 
measured in \cite{JHEP} with sufficient accuracy, so we were interested in
the effect of a small chemical potential, taken as imaginary to preserve 
positivity of the weights. The difference with the Potts case is that
the reweighting factors,
\be
\rho(U,\mu_2,\mu_1) \equiv \frac{\det^{N_f/4} \Dslash(U,\mu_2)}{\det^{N_f/4} \Dslash(U,\mu_1)} \quad ,
\ee
for each configuration $\{U\}$ were not computed exactly, but only estimated as
\be
\rho(U,\mu_2,\mu_1) = \left\langle \exp\left(-|\Dslash^{-N_f/8}(U,\mu_2) \Dslash^{+N_f/8}(U,\mu_1) \eta|^2 + |\eta|^2 \right) \right\rangle_\eta \quad ,
\ee
where $\eta$ is a Gaussian random vector. Since $N_f=3$ in our case, the 
fractional powers of the Dirac operator were approximated to high precision
by a ratio of polynomials. 
To reduce the variance, we \linebreak actually formed uncorrelated
estimators for $\sqrt{\rho(U,\mu_2,\mu_1)}$ and multiplied them together.
Further, $\rho(U,\mu_2,0), \mu_2 > \mu_1$ was constructed as 
$\rho(U,\mu_2,\mu_1) \times \rho(U,\mu_1,0)$.
In this way, highly correlated reweighting factors were obtained for
6 values of $\mu=i\mu_I$, with $a \mu_I$ ranging from $0.01$ to $0.1$.

\begin{figure}[t!]
\vspace*{-0.5cm}
\centerline{
\scalebox{0.60}{\includegraphics{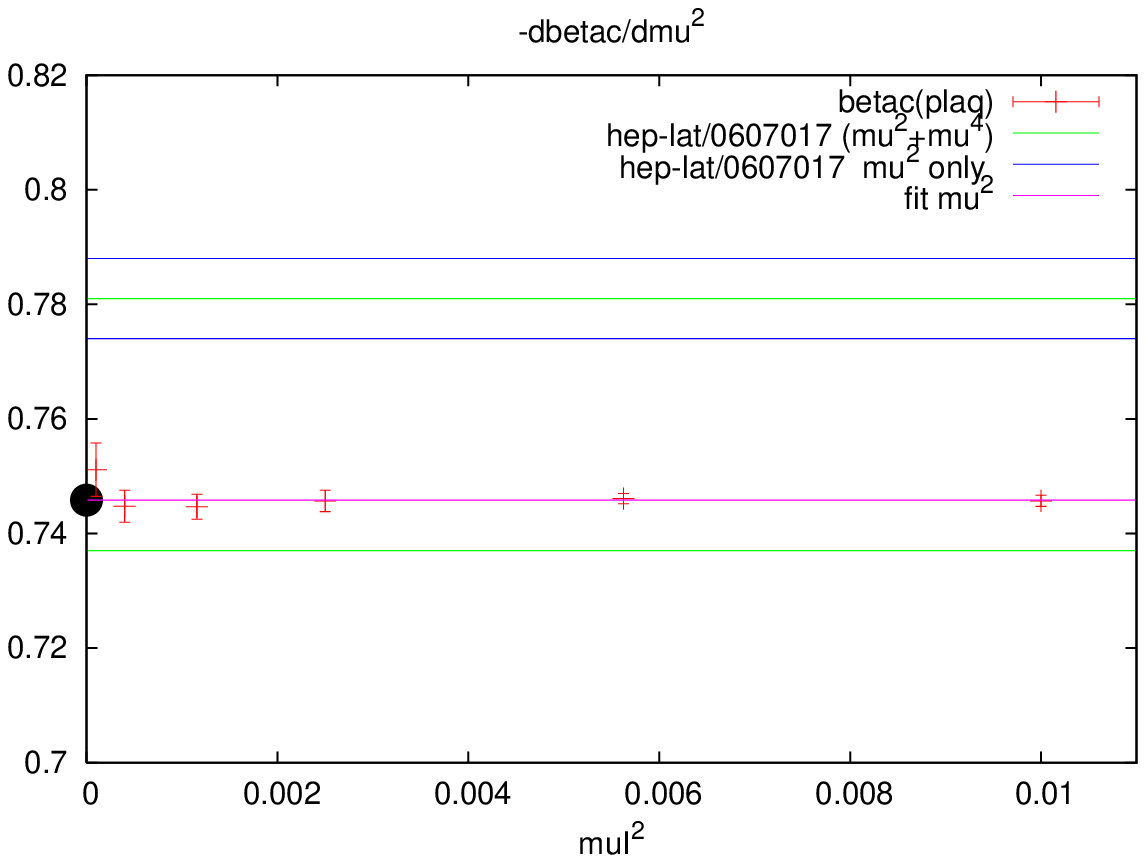}}
\hspace*{0.5cm}
\scalebox{0.60}{\includegraphics{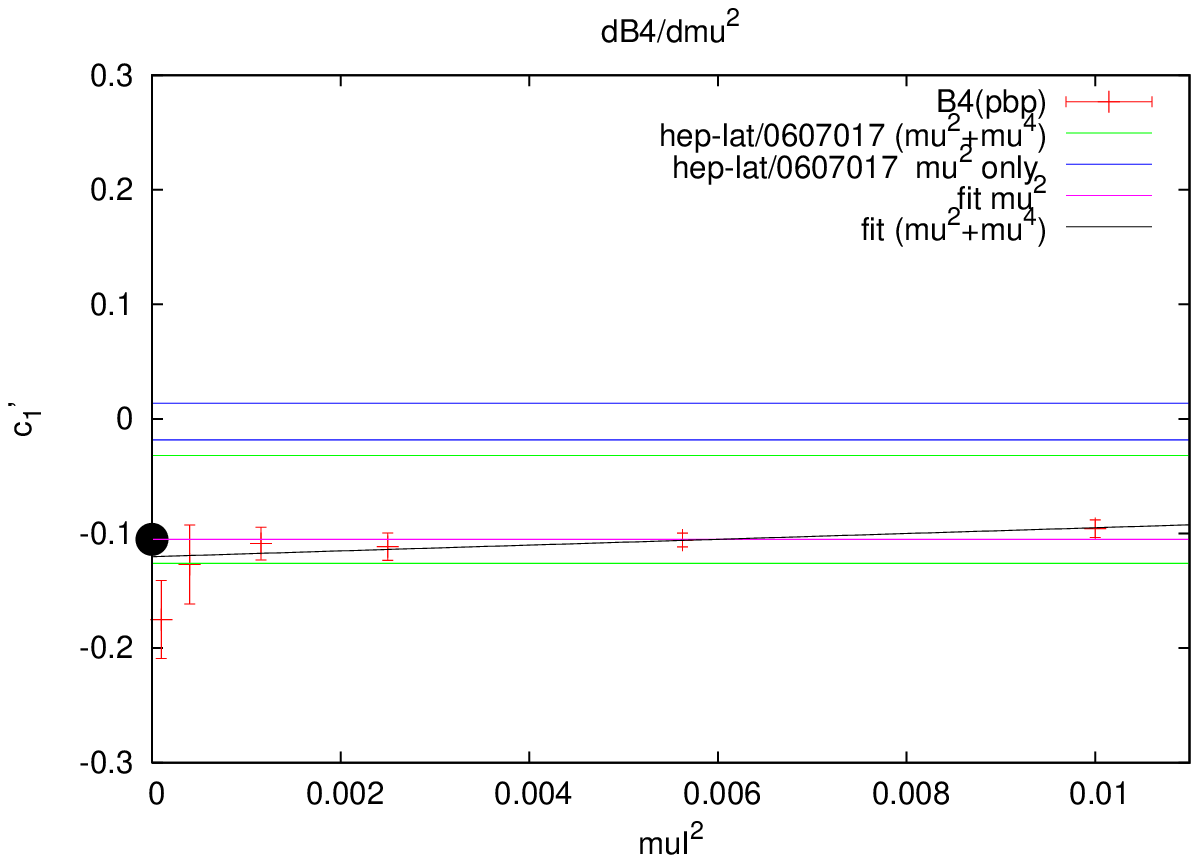}}
}
\caption{
Upon introducing a small imaginary chemical potential $\mu_I$, the 
pseudo-critical coupling $\beta_c$ and the Binder cumulant $B_4(\bar\psi \psi)$ 
vary slightly from their $\mu_I=0$ values. 
The change $\Delta{\cal O}/\Delta\mu^2$ is shown as 
a function of $(a \mu_I)^2$, for ${\cal O}=\beta_c$ ({\em left}) and for 
${\cal O}=B_4$ ({\em right}). The error bands correspond to the leading-order (blue)
and subleading-order (green) fits from Ref.\cite{JHEP}.
}
\label{grid}
\end{figure}

\begin{figure}[h!]
\vspace*{-0.2cm}
\centerline{
\scalebox{0.60}{\includegraphics{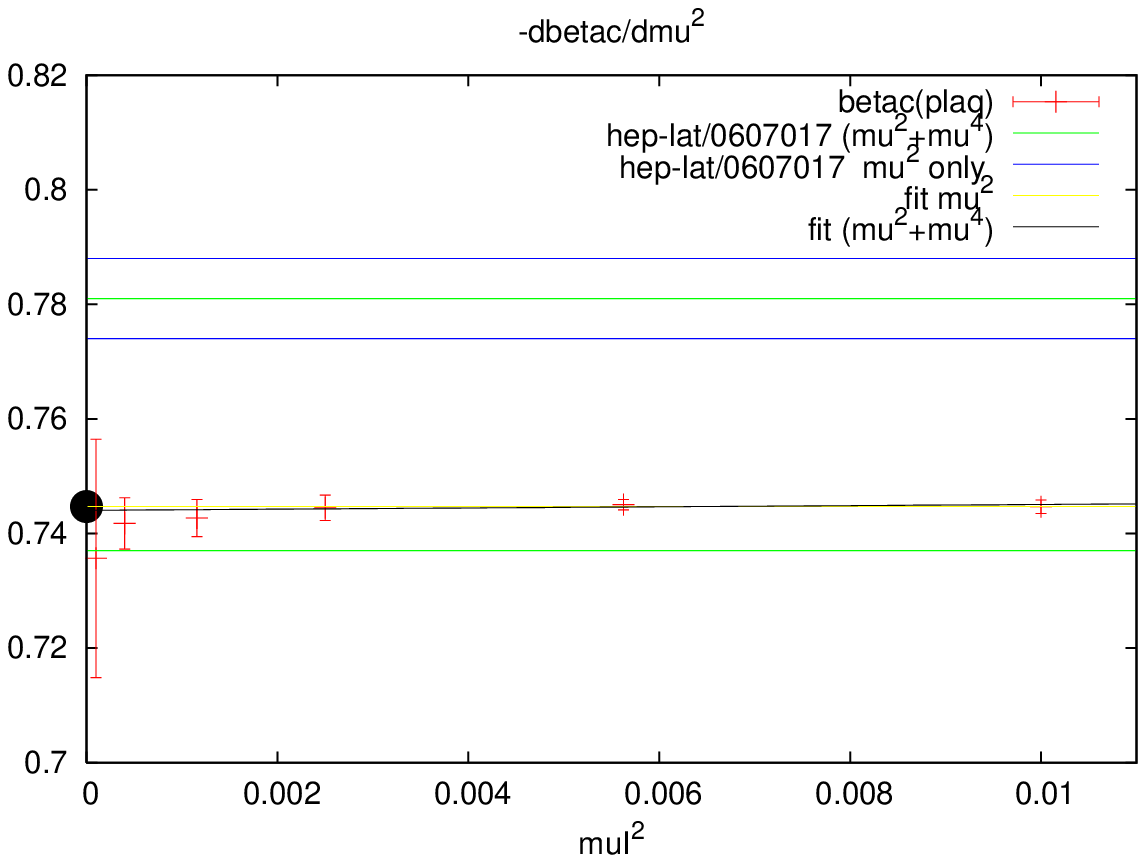}}
\hspace*{0.5cm}
\scalebox{0.60}{\includegraphics{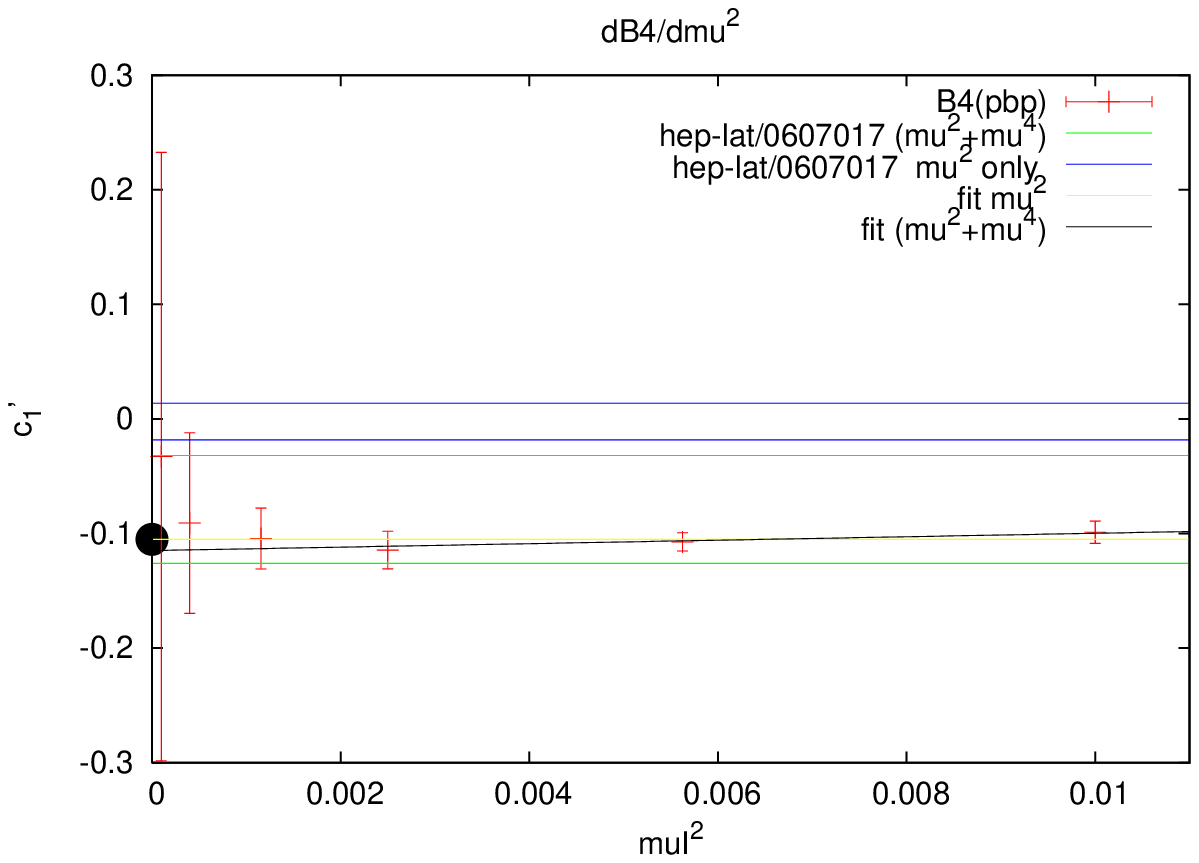}}
}
\caption{
Check of possible systematic error.
Same as Fig.~\protect\ref{grid}, but the reweighting factors for each configuration
have been multiplied by a random number uniformly distributed in $]0:2[$.   
This unbiased noise, similar to the noise in the reweighting factors 
themselves, does not bias the result.
}
\label{grid+noise}
\vspace*{-0.2cm}
\end{figure}

Crucially, the noise in the reweighting factors does not prevent the standard
application of reweighting, because these factors enter {\em linearly} in the
numerator and denominator of the reweighted expectation value
$\langle W \rangle = \frac{\sum_i \rho_i W_i}{\sum_i \rho_i}$.
Difficulties arise only when one has to form $\langle W \rangle^k$, because
the same stochastic estimator for configuration $i$ is used $k$ times and
the statistical error, when taken to an even power, introduces a bias.
This is a relevant concern, since we want to measure the $4^{th}$ cumulant of
$\bar\psi \psi$. However, this bias is of order ${\cal O}(1/N^{k-1})$ and 
disappears as the Monte Carlo sample size $N$ grows. 
To test for such bias, we magnify
it by multiplying every estimated weight by a random number drawn uniformly
in $]0,2[$. The results of the analysis, for the derivative $c_{10}$ of the
pseudo-critical coupling eq.(\ref{eq:betac_QCD}) and for the coefficient $c_1'$ 
in the Binder cumulant expansion eq.(\ref{eq:B4_QCD}), are presented Fig.~7.
They are to be compared with Fig.~6, where the original weights were used.
No bias is visible in Fig.~7, leading us to conclude that no bias is present
in Fig.~6 either.
Note that our sample size is very large: we analyzed about 5 million 
$8^3\times 4$ configurations at 21 $\beta$-values.
This large statistics was made possible by using the Grid at CERN.
The actual running time was less than two weeks.

\subsection{Comparing methods A and C}

The final results from method C (Fig.~6) are
\be
c_{10} = 0.746(1)  
~~~~~~~~~~~~ {\rm or} ~~~~~~~~~~~~~~~~~~
\frac{T_0(m_0^c,\mu)}{T_0(m_0^c,0)} 
= 1 - 0.637(1) \left( \frac{\mu}{\pi T_0} \right)^2 + \cdots
\ee
for the pseudo-critical temperature, using the two-loop $\beta$-function
to convert to physical units, and
\be
c_1' = -0.105(15)
~~~~~~~~~~~~ {\rm or} ~~~~~~~~~~~~~~~~~~
\frac{m_c(\mu)}{m_c(0)} = 1 - 3.3(5) \left( \frac{\mu}{\pi T} \right)^2 + \cdots
\ee
for the curvature of the critical surface,
where the value of $b_{10}$ eq.(\ref{eq:B4_QCD}) was taken from \cite{JHEP}:
$b_{10} \approx 13.6$. 
A subleading term would be visible as a slope in the fits Fig.~6. 
Indeed, a small effect is visible on the right, corresponding to
$c_2'= -2.5 \pm 1.2 $.

These results are consistent with those of method A, provided subleading
terms are included in fitting the latter (green error bands in Fig.~6), even
if they are statistically almost unconstrained. This illustrates the 
difficulty of estimating the systematic error of truncating the Taylor
expansion used in fitting. This difficulty is eliminated in our new method,
which in addition is about 100 times more efficient.

However, method A gathers statistics over a broader range of chemical
potentials, and with updated statistics exceeding 15 million configurations,
now allows us to clearly identify subleading Taylor terms. 
They are visible from the curvature of the fits of the Binder 
cumulant as a function of the quark mass Fig.~8 (left). The complete 
ansatz eq.(\ref{eq:B4_QCD}) was used, 
now yielding $c_1'=-0.074(28)$ and $c_2'=-1.0(5)$
with a $\chi^2$ of 23 for 21 d.o.f.
Note the consistency of methods A and C, also for the subleading order
whose sign contributes to further shrinking the region of first-order 
transitions.

Fig.~8 (right) shows the same results, after subtraction of the fitted 
mass dependence, as a function of $(a \mu_I)^2$.
The fit is shown by the lower parabola.
Now, the results of Fig.~6 (right) (leading and subleading terms) are shown
in the same figure as the upper parabola. The agreement between the
two independent methods is remarkable, given that method C only probes
the region $(a \mu_I)^2 \leq 0.01$ where agreement is near-perfect.

\begin{figure}[t!]
\vspace*{-0.3cm}
\centerline{
\scalebox{0.60}{\includegraphics{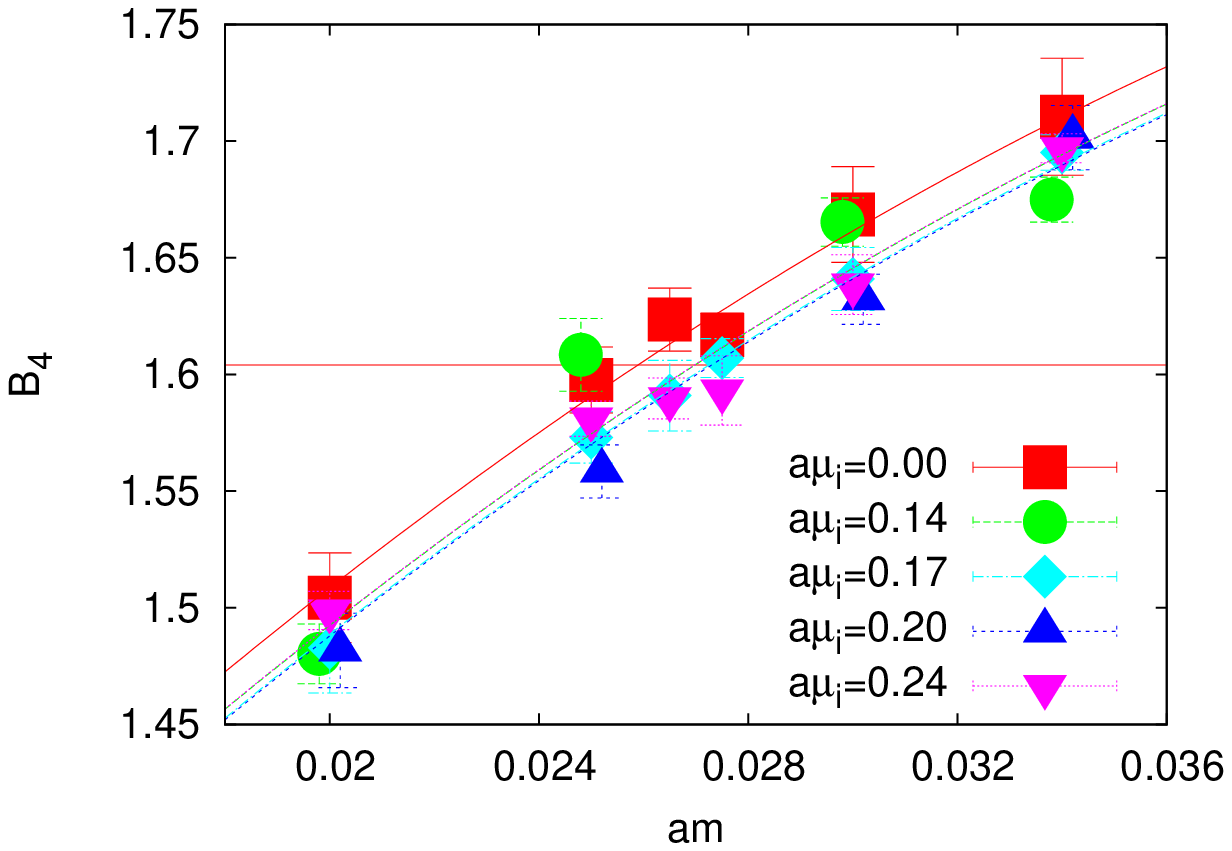}}
\hspace*{0.5cm}
\scalebox{0.60}{\includegraphics{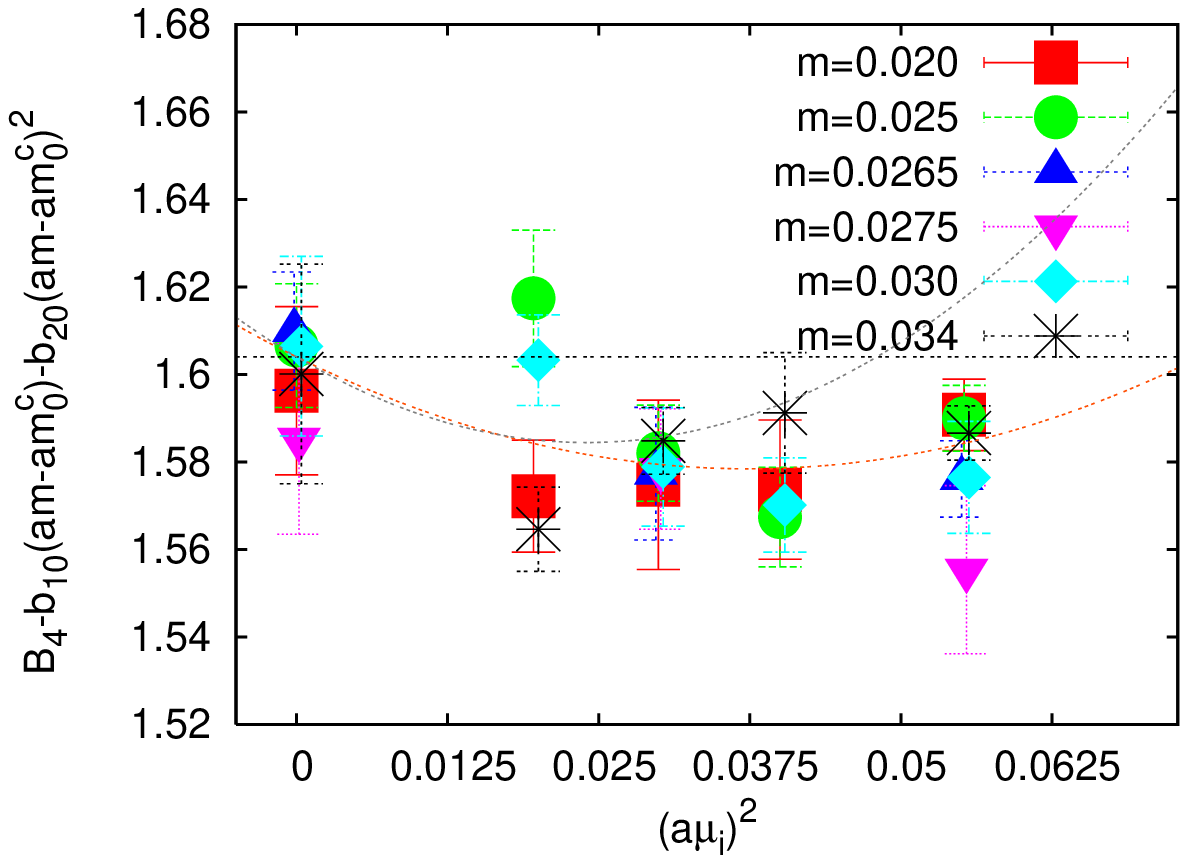}}
}
\caption{
Direct measurements of the Binder cumulant ({\em left}), updated from
Ref.~\protect\cite{JHEP}. Subleading dependence on $m$ becomes visible.
Comparison with the reweighting approach as a function of $\mu^2$ ({\em right})
shows remarkable agreement, both in the leading and subleading terms.
}
\label{global}
\end{figure}

\subsection{Towards the continuum limit}
\label{sec:Nt6}

Of course, cutoff errors on our $N_t=4$ ($a\sim 0.3$fm) lattice can be large,
and it is essential to perform a continuum extrapolation. To this end, we 
are pursuing our project on $N_t=6$ lattices, and present some preliminary
results Fig.~9.

The left figure illustrates cutoff effects on the critical bare quark mass,
$m_0^c$, corresponding to a second-order transition at $\mu=0$ in $N_f=3$ QCD.
One can see that the quark mass, expressed in units of the temperature,
must be reduced by a factor $\sim 5$ on $N_t=6$ lattices. A similarly large
effect is present in the resulting pion mass, $m_\pi^c$, measured at zero
temperature for quark mass $m_0^c$. The ratio $m_\pi^c/T_c$ decreases from
$1.680(4)$ ($N_t=4$) to $0.954(12)$ ($N_t=6$), so that a naive $a^2$ extrapolation
would give $\sim 0.4$ in the continuum!
This very large cutoff effect is consistent with earlier 
indications~\cite{Karsch_improved,MILC} and with a new study~\cite{Fodor_Nt6},
all suggesting that the transition becomes much weaker in the continuum limit.
Note that the cutoff effect on the hadron spectrum is comparatively mild,
so that the net effect of a finer lattice is to dramatically push the
critical surface Fig.~2 toward the origin, while leaving the physical point
untouched. 
Thus the gap between the critical surface and the physical point widens, 
pushing the critical point in Fig.~2 (left) to larger values of $\mu_E$.

Our second preliminary result, Fig.~9 (right), shows the curvature of the
pseudo-critical coupling $d\beta_c/d(a\mu)^2|_{\mu=0}$ for the critical
quark mass $m_0^c$. The error band corresponds to the $N_t=4$ study.
The trend is for $d\beta_c/d(a\mu)^2$ to be smaller for $N_t=6$,
while if we use the two-loop $\beta$-function \linebreak to convert to physical units,
one should observe $d\beta_c/d(a\mu)^2 \propto N_t^2$.
Instead of increasing by $(6/4)^2$, our measured value seems to decrease.
Now, for $N_t=4$ already, the
estimated curvature of the \linebreak pseudo-critical line $T_c(m_0^c,\mu)$ was
about 3 times
less than that of the experimental freeze-out curve \cite{Cleymans}. These two curves
appear to become more clearly separated as $a\to 0$, which also reduces $m_0^c$.

\begin{figure}[t!]
\vspace*{-0.3cm}
\centerline{
\scalebox{0.59}{\includegraphics{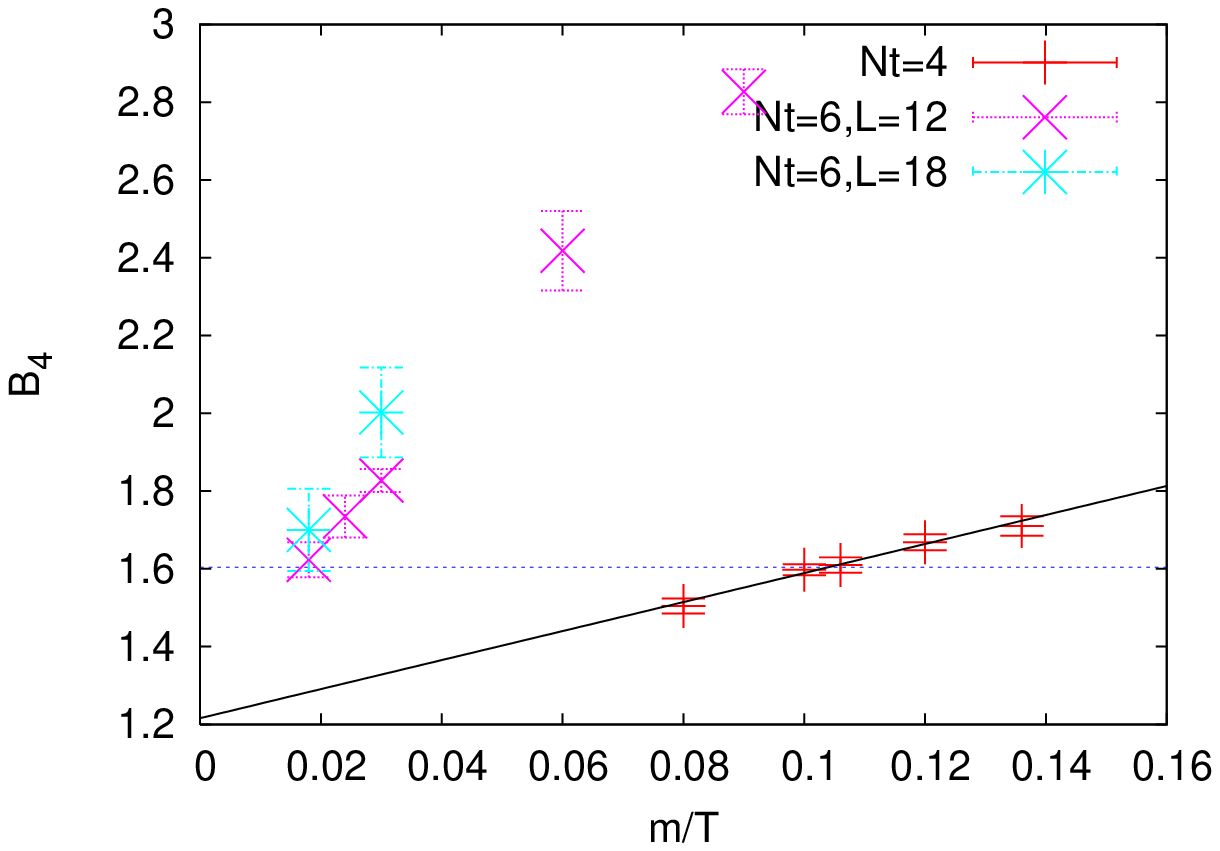}}
\hspace*{0.5cm}
\scalebox{0.60}{\includegraphics{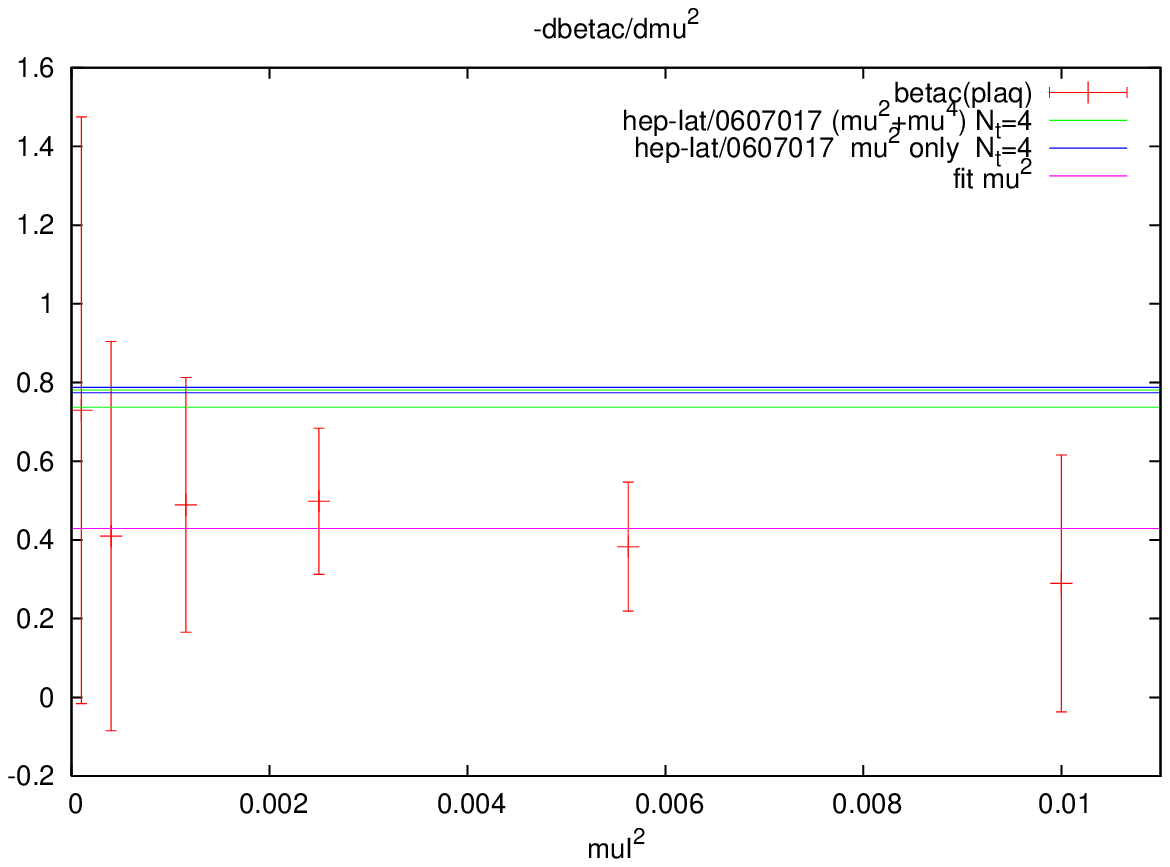}}
}
\caption{
{\em Left:} determination of the critical quark mass in the 
$N_f=3$ theory. The bare quark mass decreases (in units of $T_c$) as the
continuum limit is approached. The corresponding pion mass (measured at
$T=0$) also decreases.
{\em Right:} preliminary result for the curvature of the pseudo-critical line
for $N_t=6$. In physical units, the curvature is {\em smaller} than for
$N_t=4$. 
}
\label{Nt6}
\end{figure}

\section{Discussion}

Our findings including next-to-leading terms on $N_t=4$ predict the ``exotic'' scenario 
of Fig.~2 (right): the region
of first-order transitions shrinks as the chemical potential is turned on,
so that the \linebreak chiral critical surface does not intersect the physical line, 
and there is no chiral critical point in QCD. 
This statement, which goes against conventional wisdom, gets qualified 
by a number of systematic errors. Like [4--8,10,11,14], we use
staggered fermions with the rooting trick, which is potentially unsafe for
very light quark masses. Next, the curvature of the
critical surface varies with the cutoff, and presently the rate of this
change is unknown.
It also changes from the $N_f=3$ case presented here to
the $N_f=2+1$ theory, although we found in \cite{JHEP} that for $N_t=4$ the
sign of the curvature remains unchanged. 
Finally, once $|\mu|\sim T_c$ higher order terms may become relevant.

Despite these caveats, we believe the qualitative picture to be robust.
If the continuum critical quark mass can be Taylor expanded as per eq.(\ref{eq:m_c}) 
with coefficients ${\cal O}(1)$, the critical surface in Fig.~2 rises 
``almost vertically'' no
matter the sign of its curvature, and a critical point at small $|\mu/T|$
implies a fine-tuning of the physical quark masses, so as to be very close
to the critical line at $\mu=0$. Such a fine-tuning seems {\em unnatural},
and indeed the $\mu=0$ critical line seems to recede considerably in the 
continuum limit, now requiring a large curvature of the opposite sign to
what we observe in order to accomodate a critical point at $|\mu/T| < 1$. 

Finally, the object of our study is the {\em chiral} critical surface.
Our findings do not exclude additional critical structure due to 
non-chiral physics causing the phase transitions Fig.~3, bottom right. 

\section*{Acknowledgements}
We thank Misha Stephanov for discussions.
We thank the Center for Theoretical Physics, MIT, and the Isaac Newton
Institute, Cambridge, for hospitality. We thank the Minnesota
Supercomputer Institute for computer resources.
S.K. acknowledges the Korea Research Foundation grant KRF-2006-C00020
funded by the Korean Government (MOEHRD Basic Research Promotion Fund).
The grid-related computing presented here has been performed on the
EGEE infrastructure (EGEE is a project funded by the European Union;
contract INFSO-RI-031688) using the Ganga tool ({\em http://cern.ch/ganga}).
We would like to acknowledge the support of the EGEE
application support from CERN (IT/PSS/ED) and in particular A.~Maier, 
P.~Mendez, J.T.~Moscicki, and M.~Lamanna.


\end{document}